\documentclass[final,10pt,aps,longbibliography,twocolumn,superscriptaddress]{revtex4-1}
\usepackage{amsmath,amssymb,bm}
\usepackage{graphicx,color}
\usepackage[squaren]{SIunits}
\usepackage[english]{babel}
\usepackage{amstext}
\usepackage{amsthm}
\usepackage{latexsym}
\usepackage{array}
\usepackage{color}
\usepackage{float}
\usepackage{microtype}
 
\usepackage{multirow}
\usepackage{dcolumn}
\definecolor{url}{RGB}{0,20,160}
\usepackage[colorlinks=true,linkcolor=blue,citecolor=blue,urlcolor=url]{hyperref}
\usepackage[usenames,dvipsnames,svgnaes,table]{xcolor}

\usepackage{natbib}
\makeatletter
\def\frutiger{cmss10 }
\def\frutigerbold{cmssbx10 }
=\frutigerbold at 14pt
=\frutiger at 8pt
=\frutigerbold at 12pt
=\frutigerbold at10pt
=\frutigerbold at 8pt
=\frutiger at 8pt
=\frutigerbold at 31pt
\def\@caption@tabnum@sep{\figtextfont{{ }{\bf\textbar}{ }}}%
\def\fnum@table{{\bf\tablename~\thetable}}
\renewenvironment{table}{\@float{table}\def\textbf##1{{\fignumfont ##1}}\def\bf{\fignumfont}}{\end@float}
\def\@caption@fignum@sep{\figtextfont{{ }{\bf\textbar}{ }}}%

\renewcommand{\fnum@figure}{\bf Fig. \thefigure}
\def\@startsection#1#2#3#4#5#6{%
\if@noskipsec\leavevmode\fi
\par\@tempskipa #4\relax
\@afterindenttrue
\ifdim\@tempskipa <\z@
\@tempskipa -\@tempskipa \@afterindentfalse
\fi\if@nobreak\everypar{}%
\else\addpenalty\@secpenalty\addvspace\@tempskipa\fi
\@ifstar{\@ssect{#3}{#4}{#5}{#6}}{\@dblarg{\@sect{#1}{#2}{#3}{#4}{#5}{#6}}}}
\def\@sect#1#2#3#4#5#6[#7]#8{%
\ifnum #2>0
\let\@svsec\@empty
\else\refstepcounter{#1}\protected@edef\@svsec{\@seccntformat{#1}\relax}\fi
\@tempskipa #5\relax
\ifdim\@tempskipa>\z@
\begingroup#6{\@hangfrom{\hskip #3\relax\@svsec}%
\interlinepenalty \@M #8\@@par}\endgroup
\csname #1mark\endcsname{#7}%
\addcontentsline{toc}{#1}{%
\ifnum #2>\c@secnumdepth\else
\protect\numberline{\csname the#1\endcsname}\fi #7}%
\else\def\@svsechd{#6{\hskip #3\relax
\@svsec #8\ifnum#2=2.\fi}%
\csname #1mark\endcsname{#7}%
\addcontentsline{toc}{#1}{%
\ifnum #2>\c@secnumdepth \else
\protect\numberline{\csname the#1\endcsname}\fi #7}}%
\fi\@xsect{#5}}
\renewcommand\section{\@startsection {section}{1}{\z@}%
{-10pt \@plus -1ex \@minus -.2ex}{.5ex }{\normalfont\Large\bfseries\sectionfont}}
\renewcommand\subsection{\@startsection{subsection}{2}{\z@}%
{10pt\@plus 1ex \@minus.2ex}{-0.5ex \@plus.2ex}{\normalfont\large\bfseries\subsectionfont}}
\def\frontmatter@title@format{\titlefont\centering}%
\def\frontmatter@title@below{\addvspace{-5pt}}%

\newcommand*\bib@heading{%
  \section{\refname}
  \fontsize{8}{10}\selectfont
}
\newcommand*\@openbib@code{%
      \advance\leftmargin\bibindent
      \itemindent -\bibindent
      \listparindent \itemindent
      \parsep \z@
}%
\newdimen\bibindent
\makeatother

\newcommand{\lbnl}{Energy Technologies Area, Lawrence Berkeley National Laboratory, Berkeley, CA 94720, USA}
\newcommand{\northwestern}{Department of Materials Science \& Engineering, Northwestern University, Evanston, IL 60208, USA}
\newcommand{\yale}{Department of Applied Physics, Yale University, New Haven, CT 06511, USA}
\newcommand{\esi}{Energy Sciences Institute, Yale University, West Haven, CT 06516, USA}

\begin{document}

	\title{Optimal Band Structure for Thermoelectrics with Realistic Scattering and Bands}
	\author{Junsoo Park}
	\email{qkwnstn@gmail.com}
	\affiliation{\lbnl}
	\author{Yi Xia}
	\affiliation{\northwestern}
	\author{Vidvuds Ozoli\c{n}\v{s}}
	\affiliation{\yale} 	
	\affiliation{\esi} 	
	\author{Anubhav Jain}
	\email{ajain@lbl.gov}
	\affiliation{\lbnl}
	\date{\today} 
	\begin{abstract}
Understanding how to optimize electronic band structures for thermoelectrics is a topic of long-standing interest in the community. Prior models have been limited to simplified bands and/or scattering models. In this study, we apply more rigorous scattering treatments to more realistic model band structures - upward-parabolic bands that inflect to an inverted parabolic behavior - including cases of multiple bands. In contrast to common descriptors (e.g., quality factor and complexity factor), the degree to which multiple pockets improve thermoelectric performance is bounded by interband scattering and the relative shapes of the bands. We establish that extremely anisotropic `flat-and-dispersive' bands, although best-performing in theory, may not represent a promising design strategy in practice. Critically, we determine optimum bandwidth, dependent on temperature and lattice thermal conductivity, from perfect transport cutoffs that can in theory significantly boost $zT$ beyond the values attainable through intrinsic band structures alone. Our analysis should be widely useful as the thermoelectric research community eyes $zT>3$.
	\end{abstract}
	\maketitle

\section{Introduction}

Thermoelectricity enables clean electricity generation and fluid-free cooling. The ultimate goal of basic thermoelectric materials research is to design or discover materials with high figure of merit $zT$, commonly expressed as
\begin{equation}\label{eq:zt1}
zT=\frac{\alpha^{2}\sigma}{\kappa_{e}+\kappa_{\text{lat}}}T.
\end{equation}
Here, the thermoelectric power factor (PF) is the product of Ohmic charge conductivity ($\sigma$) and the Seebeck coefficient ($\alpha$) squared. The total thermal conductivity $\kappa$ is the sum of electronic thermal conductivity ($\kappa_{e}$) and lattice thermal conductivity ($\kappa_{\text{lat}}$). A major challenge in achieving high $zT$ and PF is that the electronic transport quantities are linked by a set of anti-complementary correlations: \cite{complex,newandold,intuition,perspectivesonthermoelectrics,thermoelectricmaterials,compromisesynergy,advancesinthermoelectrics,advancesinthermoelectricmaterials,onthetuning,computationalthermoelectrics,computationalenergymaterials} $\sigma$ and $\kappa_{e}$ are positively correlated whereas $\sigma$ and $\alpha$ are negatively correlated. Only $\kappa_{\text{lat}}$, a lattice property, is relatively independent, though it too exhibits some positive correlation with $\sigma$ through structural symmetry. These interrelations make it difficult to determine the effect of various design strategies to optimize $zT$.

Equations based on the single parabolic band (SPB) model often underpin intuition about thermoelectric behavior. However, they tacitly assume that there is always enough (infinite) dispersion in all directions to cover the entire energy range relevant to thermoelectric phenomena. Instead, in most cases of practical interest, a band's dispersion changes in curvature (e.g., from positive to negative), crosses the Brillouin zone (BZ) boundary orthogonally, and tops out at some maximum energy. In addition to band shape considerations, thermoelectric properties can widely vary depending on what is assumed of the scattering behavior. Typical models and descriptors assume a behavior that is dominated by intraband/intravalley, elastic acoustic phonon scattering, and can be derailed when other scattering mechanisms and interband/intervalley transitions have large effects \cite{wangsnyderbook,valleytronics1,valleytronics2,roleofscattering,bandalignmentscattering}. Several studies have analytically investigated thermoelectricity using model band structures and scattering \cite{roleofscattering,bandalignmentscattering,simplescattering,optimalbandwidth,bestbandstructure}, but they had one or more of the following limitations: 1) the bands were purely parabolic or parabolic-like with infinite dispersion; 2) only a single isotropic band was considered;  3) models for scattering and/or transport were based on constant lifetimes, constant mean free paths, or at best scattering proportional to the density of states (DOS).

\begin{figure*}[tp]
\centering
\includegraphics[width=1 \linewidth]{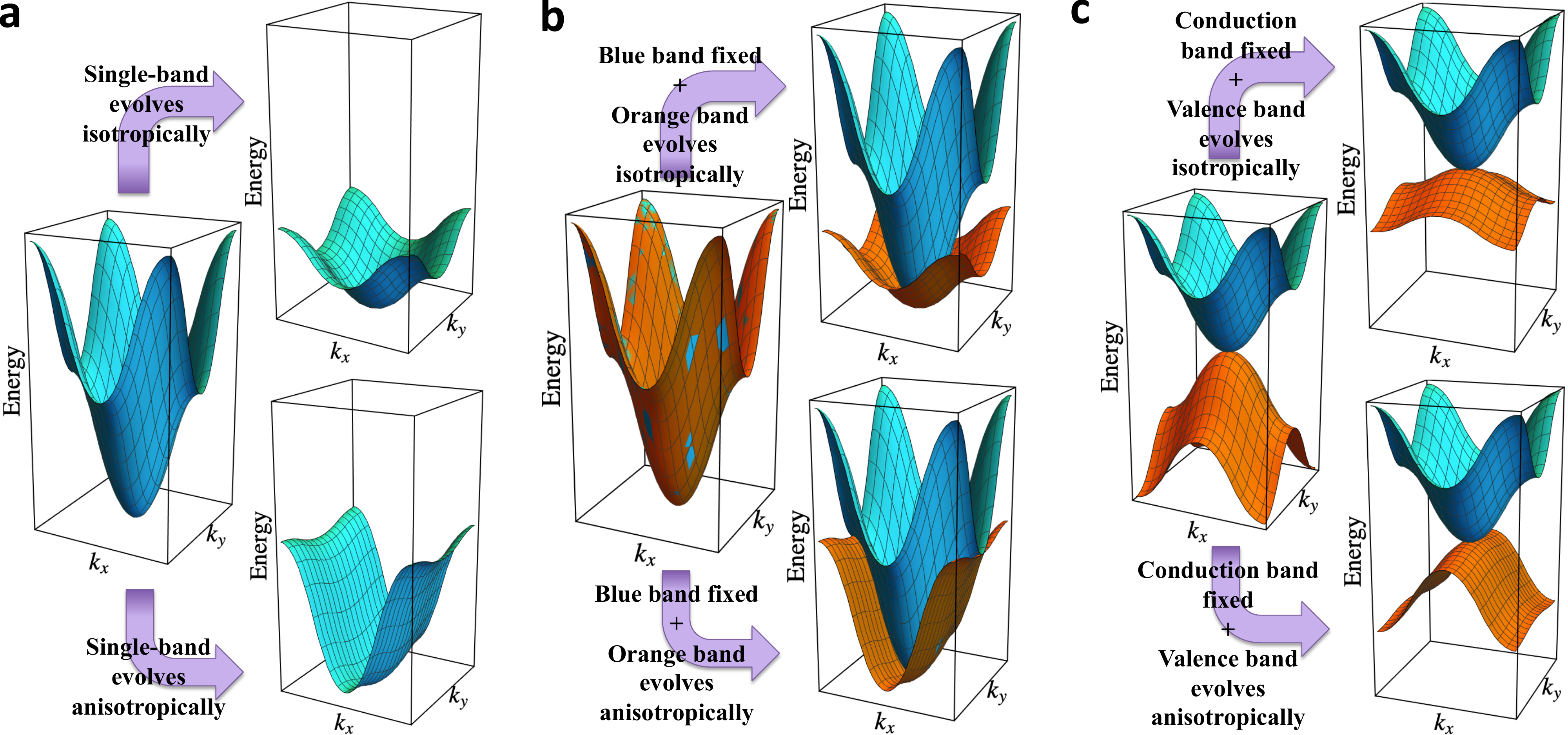}
\caption{The evolution of band structure models used in this study. \textbf{a)} A single band changes in effective mass. \textbf{b)} One band (orange) changes in effective mass while another band (blue) is fixed. \textbf{c)} The valence band (orange) changes in effective mass while the conduction band (blue) is fixed. Note that two-dimensional (2D) band structures are shown for graphical purposes. In the study where three-dimensional (3D) bands are used, two types of anisotropic evolution are considered: one where a band grows heavy in one direction and another where the band grows heavy in two directions. Each band is an upward paraboloid smoothly inflecting to an inverted paraboloid halfway to the BZ boundary.} 
\label{fig:bandevolution}
\end{figure*}

To more generally addresses the topic of optimal band structure, we create more realistic model solid-state band structures and more faithfully model carrier scattering due to multiple sources.  Our band structures are properly confined to a finite BZ with smooth inversion of upward (downward) parabolicity to downward (upward) parabolicity for describing conduction (valence) states - a key for retaining generality, physicality, and approximate compatibility with established scattering formalism. We modify established formulae for various scattering mechanisms - deformation-potential scattering (DPS), polar-optical scattering (POS), and ionized-impurity scattering (IIS) - as to capture the effects of inverted parabolicity, anisotropy, and band multiplicity on carrier lifetimes.  Refer to Methods for further details. We monitor how thermoelectric properties of one or more bands respond to variations in band shapes (see Fig. \ref{fig:bandevolution}). Our study fine-tunes conclusions drawn from simpler models on design strategies such as anisotropy, band multiplicity, and resonance levels. Finally, we determine the optimum bandwidths as a function of temperature and $\kappa_{\text{lat}}$, which improves $zT$ beyond what is normally accessible.

We start by rewriting Eq. \ref{eq:zt1} to better reflect fundamental transport relations:
\begin{equation}\label{eq:zt2}
zT=\frac{(\zeta^{2}/\sigma)}{\kappa_{e}+\kappa_{\text{lat}}}T.
\end{equation}
In Eq. \ref{eq:zt2}, the key role is played by $\zeta$, a quantity for which there appears to be no conventional name. We refer to it as the `thermoelectric conductivity'; in the Onsager-Callen formulation of coupled charge-and-heat conduction \cite{onsager1,onsager2,callen}, $\zeta$ is the quantity responsible for the thermal-gradient-to-charge-current conversion ($\mathbf{J}_{c}=\sigma\mathbf{E}-\zeta\nabla T$). That is, $\zeta$ represents the charge conductivity due to thermal driving force, the essence of thermoelectricity. Eq. \ref{eq:zt2} lifts the hidden coupling between $\alpha$ and $\sigma$ ($\alpha=\zeta/\sigma$) and correctly identifies $\zeta$ as the quantity that must be high but that $\sigma$ must be \textit{low}. That is, we desire high thermoelectric conductivity, not Ohmic conductivity - a correction to the routine but ambiguous thermoelectric adage that `electrical conductivity' must be high.

\begin{figure*}[tp]
\centering
\includegraphics[width=1 \linewidth]{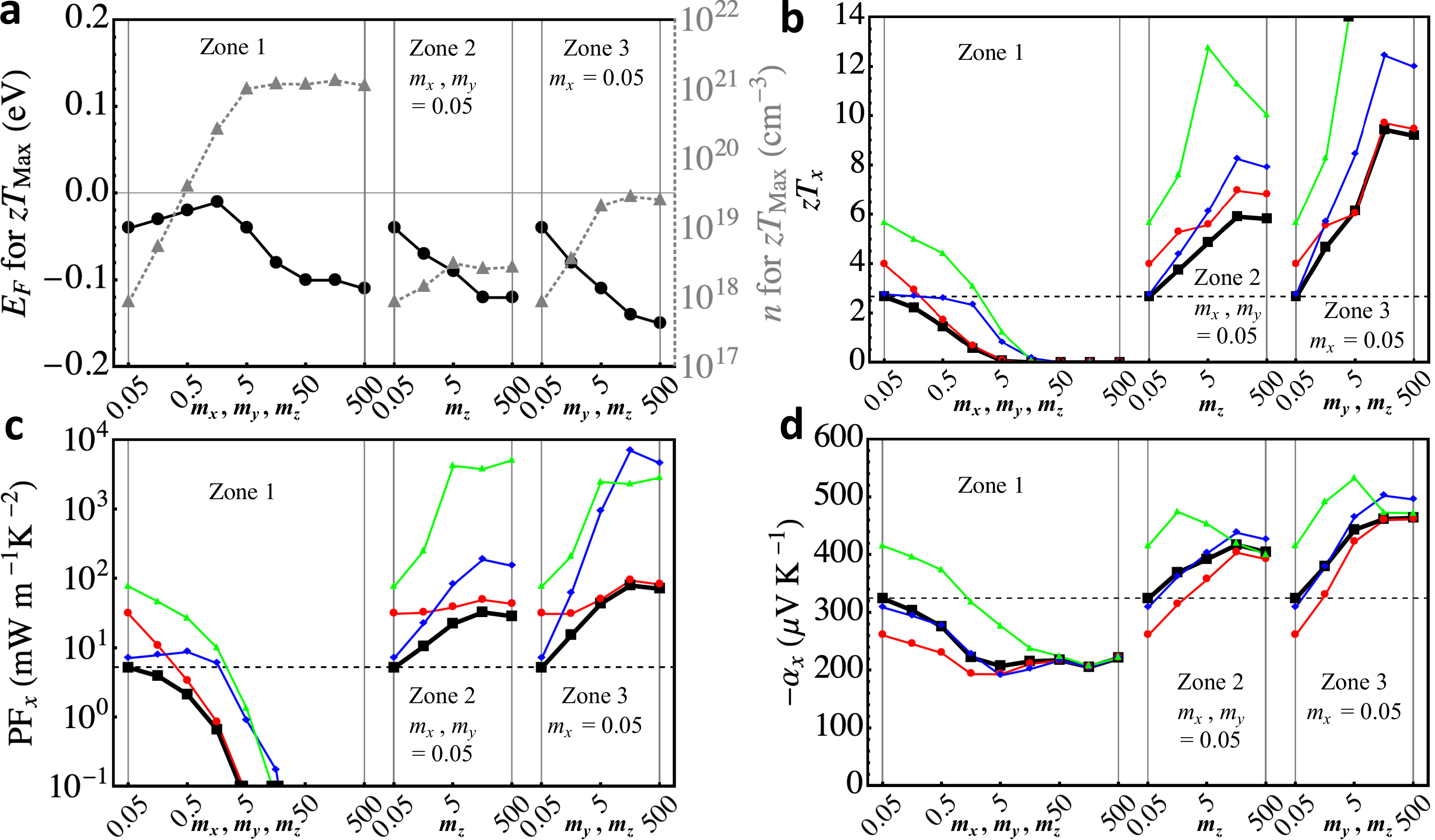}
\caption{Single-band thermoelectric properties  in the light direction ($x$) with $m_{x}=0.05$  with respect its effective mass profile. \textbf{a)} Fermi level and carrier concentrations at optimum $zT$, \textbf{b)} optimum $zT$, \textbf{c)} the power factor, \textbf{d)} and the Seebeck coefficient, in the $x$-direction. Each zone (as enclosed by vertical gray lines) indicates certain characteristic evolution: isotropic increase in $m$ from 0.05 to 500 in Zone 1, anisotropic increase in $m_{y}$ from 0.05 to 500 in Zone 2, anisotropic increase in both $m_{y}$ and $m_{z}$ in Zone 3 from 0.05 to 500. Four different scattering regimes are considered: the POS limit (\color{blue}blue\color{black}), the IIS limit (\color{green}green\color{black}), the DPS limit (\color{red}red\color{black}), and the overall effect (black). Supplementary results for $\sigma, \mu, \zeta, \kappa_{e}, L, z_{e}T$ are in Supplementary Fig. 8.} 
\label{fig:3dsingleopt}
\end{figure*}

Insights into maximizing $zT$ are attained by examining Eq. \ref{eq:zt2} through Boltzmann transport formalism \cite{dovertheoretical,ziman,btecoefficients,boltztrap},
\begin{equation}\label{eq:sigma}
\sigma=\frac{1}{V} \int \Sigma(E)\left(-\frac{\partial f}{\partial E}\right)dE,
\end{equation}
\begin{equation}\label{eq:zeta}
\zeta=\frac{1}{V T} \int (E_{\text{F}}-E)\Sigma(E)\left(-\frac{\partial f}{\partial E}\right)dE,
\end{equation}
\begin{equation}\label{eq:kappae}
\kappa_{e}=\frac{1}{V T} \int (E_{\text{F}}-E)^{2}\Sigma(E)\left(-\frac{\partial f}{\partial E}\right)dE-\frac{\zeta^{2}}{\sigma} T,
\end{equation}
where $V$ is the cell volume, $E_{\text{F}}$ is the Fermi level, $f(E)$ is the Fermi-Dirac distribution, and $\Sigma(E)=v^{2}(E)\tau(E)D(E)$ is the spectral conductivity, composed of group velocity ($v$), lifetime ($\tau$), and DOS ($D$).
The three integrands share in common the term $\Sigma(E)\left(-\frac{\partial f}{\partial E}\right)$, the source of the positive correlations between $\sigma$, $\zeta$, and $\kappa_{e}$. The integrands differ only in the power relation $(E_{\text{F}}-E)^{p}$ as $p=0, 1, 2$. This juxtaposition states that, in relative terms, low energy carriers contribute most to $\sigma$, high-energy carriers contribute most to $\kappa_{e}$, while it is the medium-energy carriers that are most responsible for $\zeta$. That is to say, if one wished to increase $\zeta$ \textit{relative to} $\sigma$ and $\kappa_{e}$, then $\Sigma(E)$ should be high in some medium-energy range and low elsewhere. The results that follow are interpreted with this picture in mind.

\section{Results}

\begin{figure*}[tp]
\centering
\includegraphics[width=1 \linewidth]{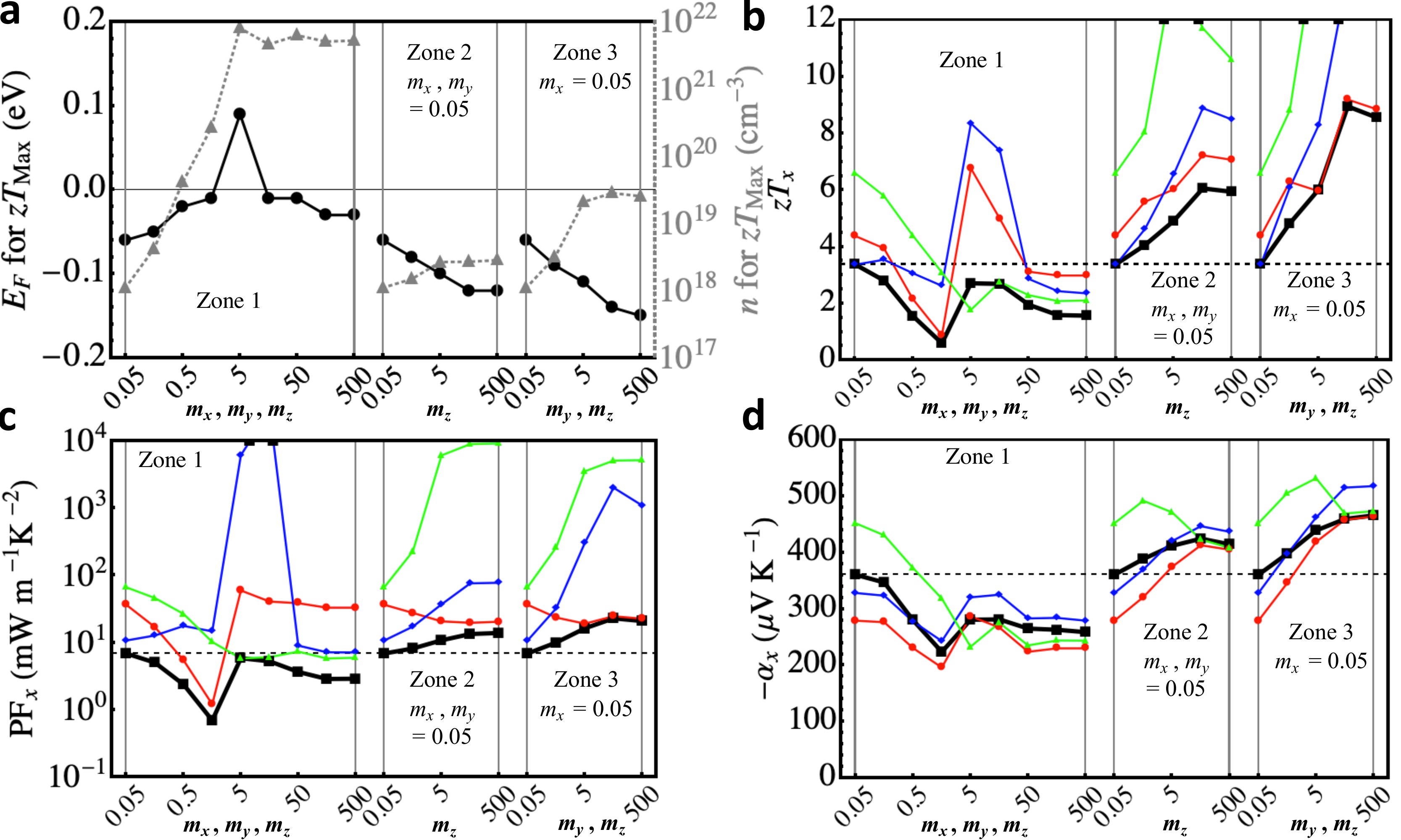}
\caption{Two-band thermoelectric properties in the light direction ($x$) with respect to the evolution the second band's effective mass profile while the first band is fixed at $m_{x}=m_{y}=m_{z}=0.05$ with $s_{\text{int}}=0.5$. \textbf{a)} Fermi level and carrier concentrations for optimum $zT$, \textbf{b)} optimum $zT$, \textbf{c)} the power factor, \textbf{d)} and the Seebeck coefficient, in the $x$-direction. Each zone (as enclosed by vertical gray lines) indicates certain characteristic evolution of the second band: isotropic increase in $m$ from 0.05 to 500 in Zone 1, anisotropic increase in $m_{y}$ from 0.05 to 500 in Zone 2, anisotropic increase in both $m_{y}$ and $m_{z}$ in Zone 3 from 0.05 to 500. Four different scattering regimes are considered: the POS limit (\color{blue}blue\color{black}), the IIS limit (\color{green}green\color{black}), the DPS limit (\color{red}red\color{black}), and the overall effect (black). Supplementary results $\sigma, \mu, \zeta, \kappa_{e}, L, z_{e}T$ are in Supplementary Fig. 10.} 
\label{fig:3ddoubleopt}
\end{figure*}

\subsection{Optimal Performance - Single Band} 

Here we investigate how a single band may yield the highest $zT$ with $E_{\text{F}}$ optimized for it. The performance is evaluated for different band structure shapes as depicted in the three `zones' of Fig. \ref{fig:3dsingleopt}: 1) isotropic increase in $m$, 2) anisotropic increase in $m$ in one direction (`unidirectional anisotropy'), and 3) anisotropic increase in $m$ in two directions (`bidirectional anisotropy'). The Seebeck coefficient predicted at a fixed $E_{\text{F}}$ is provided in Supplementary Discussion, which pinpoints how and why our model predictions deviate from the SPB model. The fluctuations in optimal $E_{\text{F}}$, displayed in Fig. \ref{fig:3dsingleopt}a, is also analyzed there.

We first consider the case where $m$ varies isotropically (Zone 1 in Figs. \ref{fig:3dsingleopt}b--d). As expected, a light band is definitely preferred: the PF and $zT$ both decrease with increase in $m$, as numerous studies agree upon \cite{loweffmass,materialdescriptors,complexity}. A lighter band has higher mobility ($\mu$) and thus is less needy of carrier concentration ($n$) in providing a given value of $\sigma$ ($\sigma=n\mu$), which helps retain high $\alpha$.

We observe that anisotropy is immensely beneficial (see Zones 2 and 3 in Fig. \ref{fig:3dsingleopt}). Because DPS is almost exactly proportional to DOS, the performance under DPS is a clear indicator of the important role played by the energy-dependence of group velocity, $\langle v^{2}(E) \rangle$, which steepens with band anisotropy to enhance performance. See Supplementary Fig. 6  for the schematic. Steepening $\langle v^{2}(E) \rangle$ increases $\zeta$ over $\sigma$, simultaneously lowering optimal $E_{\text{F}}$. We make three major observations. First, in terms of $zT$, bidirectional anisotropy (one light, two heavy directions) outperforms unidirectional anisotropy (two light, one heavy direction). This is because $\langle v^{2}(E) \rangle$ in the former evolves to a one-dimensional-like profile, which is steeper than the two-dimensional-like profile that $\langle v^{2}(E) \rangle$ evolves to in the latter. Second, toward the extreme limit, both types of anisotropy plateau in performance. This occurs for two reasons: for one, $\langle v^{2}(E) \rangle$ converges to the respective low-dimensional linear limits, and for two, extreme anisotropy exhausts `low-energy voids'. Refer to Supplementary Discussion for details. Third, because IIS and POS are less dependent on $D(E)$ than DPS, anisotropy is even more beneficial when they are the dominant mechanisms. Eventually though, because DPS increases most rapidly with DOS, it becomes dominant as anisotropy grows large. Overall, we observe that anisotropy improves $zT$ by as much as a factor of 3 above the isotropic value.

Values in Fig. \ref{fig:3dsingleopt} would be lower if $m_{x}$ were larger and $\kappa_{\text{lat}}$ were higher. In Supplementary Fig. 7 , we show under $m_{x}=0.1$ and $\kappa_{\text{lat}}=1$ W m$^{-1}$ K$^{-1}$, $zT$ is limited to 5 rather than 9, in a closer neighborhood of the state-of-the-art, but draw the same relative benefit from anisotropy.

\subsection{Optimal Performance - Multiple Bands} 

Realistic band structures often feature multiple bands near the Fermi level. One of the best designs known for increasing $\sigma$ without paying a penalty on $\alpha$ is multiplicity of band pockets aligned in energy \cite{halfheuslerbanddegeneracy,zintlorbitalengineering,chalcopyritebandconvergence1,bandconvergencereview,pbtebandconvergence}. Band multiplicity comes in various forms, however; we therefore examine the effects of (i) multiplicity of identical bands, (ii) coexistence of inequivalent bands (with varying the second band shape), and (iii) bipolar transport in the presence of valence and conduction bands (with varying valence band shapes). These band structures are illustrated in Fig. \ref{fig:bandevolution}b-c.  As justified in the Methods section, our modeling of interband/intervalley scattering (henceforth inter-scattering) expands the phase space owing to the additional band and uses the factor $s_{\text{int}}=0.5$ making it half as strong as intraband/intravalley scattering (henceforth intra-scattering).  For comparison, we also provide results obtained with $s_{\text{int}}=0$ (no inter-scattering) in Supplementary Fig. 11. 

We start from two identical bands with aligned band minima, the first of which is isotropic and fixed while the second band then evolves according to Fig. \ref{fig:bandevolution}b. $E_{\text{F}}$ is again optimized for maximum $zT$. The results are plotted in Fig. \ref{fig:3ddoubleopt}. The left edge of Zone 1 for each plot, where the two bands are identical, represents symmetry-degenerate band pockets. This offers higher $zT$ and PF as compared to the case of a single band (Fig. \ref{fig:3dsingleopt}) though less than by twofold. Two identical bands result in essentially identical $\alpha$ whereas $\sigma$ draws benefits from doubled $n$ somewhat negated by inter-scattering. One question of interest is the effect of increasing the number of identical carrier pockets. It is generally known that the more pockets the better, though it is straightforward even from our simplified analysis that doubling their number does not double the PF or $zT$ due to inter-scattering. For $N_{v}$ band pockets, $n\propto N_{v}$ while $\tau\propto \left(1+s_{\text{int}}(N_{v}-1)\right)^{-1}$. Then $\sigma\propto N_{v}\left(1+s_{\text{int}}(N_{v}-1)\right)^{-1}$, which as $N_{v}$ grows saturates to $s_{\text{int}}^{-1}$. For example, with $s_{\text{int}}=0.5$ that we assume, the maximum PF gain even with an infinite number of identical pockets is a factor of 2. In fact, if inter-scattering is somehow stronger than intra-scattering, $s_{\text{int}}>1$, then $N_{v}$ is detrimental. As such, the benefit of $N_{v}$ is bounded by the degree of inter-scattering, whose minimization should be a priority of multi-band strategies, e.g., by focusing on pockets located at distant pockets in the BZ \cite{bandconvergence}. Furthermore, if $\kappa_{e}>>\kappa_{\text{lat}}$, then $N_{v}$ is rather unimportant for $zT$ because $\kappa_{e}$ increases as much as the PF. If $\kappa_{\text{lat}}=0$, hypothetically, then $N_{v}$ would have no effect as it cancels exactly for the PF and $\kappa_{e}$.

Next, keeping the principal first band fixed in shape and maintaining $s_{\text{int}}=0.5$, we make the second band heavier. As it turns heavier isotropically (Zone 1 in Fig. \ref{fig:3ddoubleopt}), $zT$ and the PF increasingly suffer until they sink well below even the values that the fixed principal band alone generates (compared to Fig. \ref{fig:3dsingleopt}). This means that non-symmetry-related, accidentally degenerate pockets harm $zT$ if their band masses in the transport direction ($m_{x}$) are sufficiently different. Two main reasons account for this. As the second band grows heavier in the transport direction ($x$), its direct contribution to transport diminishes. It also indirectly sabotages the lighter principal band by triggering heavier inter-scattering overall. This holds until the second band becomes narrow enough for it to function as a resonance level and selectively scatter low-energy carriers, whereby $zT$ and the PF rebound. They do not fully recover the values generated by the original twin degenerate bands unless DPS or POS dominates. The presence of strong IIS, due to the high impurity concentration required for doping a very heavy band, could eclipse the resonance level effect from manifesting.

If the second band evolves anisotropically in the $y$ and/or the $z$ directions, the thermoelectric response is largely similar to what is seen for a single band turning anisotropic. Anisotropy increases $\alpha$ and the PF as well as $zT$ until they plateau. Also, $zT$ is not noticeably higher here than in the case of a single anisotropic band because the anisotropic band dominates transport and $\kappa_{e}>>\kappa_{\text{lat}}$. This again is a nod to the decreasing importance of the band multiplicity if $\kappa_{e}>>\kappa_{\text{lat}}$.

Another two-band situation is a semimetallic one in which there exists a `conduction band' and a `valence band' with no gap in between, triggering bipolar transport. Bipolar effect is a significant suppressor of the Seebeck coefficients of metals and small-gap semiconductors. Extrapolating the lessons from above, it is rather straightforward that for $\zeta$ to be large in magnitude (positive or negative), $\Sigma(E)$ must be highly asymmetric about the Fermi level, juxtaposing mobile and anisotropic `conduction' bands against isotropically heavy `valence' bands or vice-versa. We confirm this by fixing the conduction band and evolving the valence band as described in Fig. \ref{fig:bandevolution}c.  The results are in Supplementary Fig. 12 and Discussion. It is therefore no surprise that high-performing semimetals and narrow-gap semiconductors feature quite drastic band asymmetries about the Fermi level \cite{cosiyi,cosisame,mos2yi,ybal3epw,thermoelectricsemimetals,asymmetricbands}.

\begin{figure*}
\includegraphics[width=1 \linewidth]{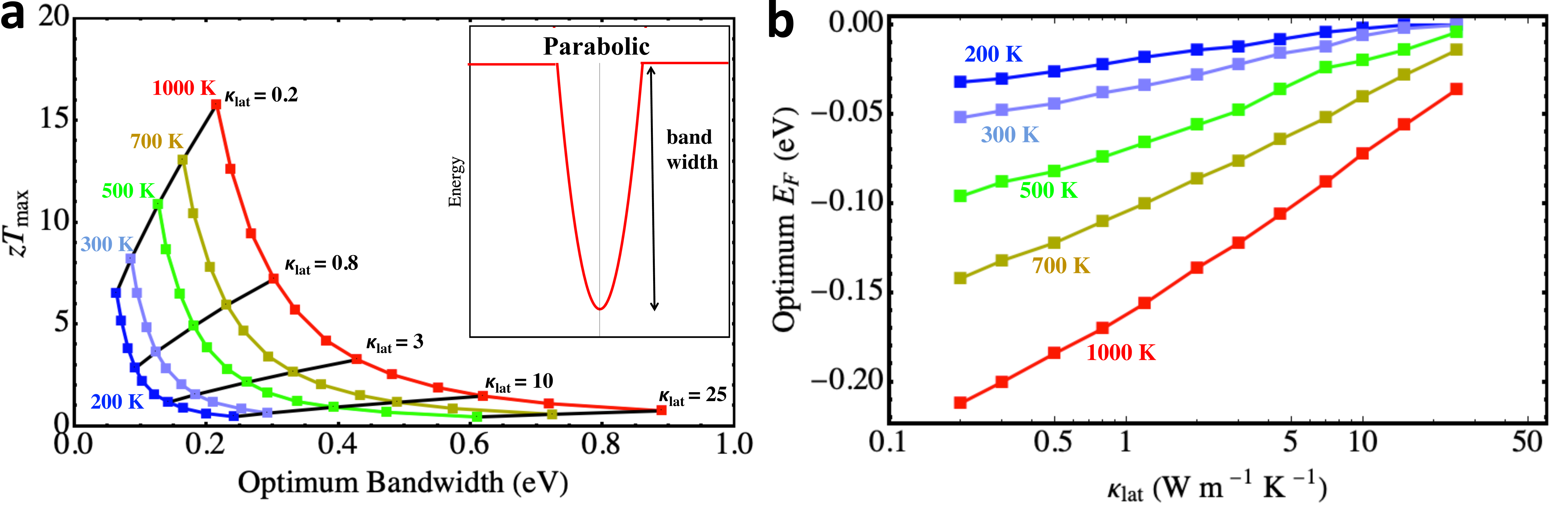}
\caption{Optimum bandwidth, Fermi level, and $zT$. \textbf{a)} $T$-and-$\kappa_{\text{lat}}$-dependent optimum bandwidth and $zT$ under DPS for an isotropic 3D parabolic band of $m_{\text{GaAs}}=0.067$, and \textbf{b)} the optimum Fermi level for each point. The lower the $\kappa_{\text{lat}}$, the lower the optimum $E_{\text{F}}$ and $W_{\text{opt}}$. $\kappa_{\text{lat}}$ is given in W m$^{-1}$K$^{-1}$.}
\label{fig:bandwidth}
\end{figure*}

\subsection{Optimum Bandwidth from Perfect Transport Cutoff} 

The relative energy ranges from which $\sigma$, $\zeta$, and $\kappa_{e}$ draw contributions imply that the best performance would be obtained by suppressing both low-energy contributions (to suppress Ohmic current) and high-energy contributions (to suppress thermal current) thus limiting transport only to a certain medium-energy range. Accordingly, we investigate the scenario in which the contribution to transport abruptly vanishes at some optimum energy (see the inset in Fig. \ref{fig:bandwidth}a), which we define as the optimum bandwidth ($W_{\text{opt}}$). It essentially represents optimum transport distribution width. Mathematically, $W_{\text{opt}}$ is obtained by solving for the following as to maximize $zT$:
\begin{equation}\label{eq:optimizezt}
W_{\text{opt}}=\text{argmax}_{W}\left[\frac{\zeta^{2}(W)/\sigma(W)}{\kappa_{e}(W)+\kappa_{\text{lat}}}T\right]
\end{equation}
where, for instance, $\sigma(W)=\frac{1}{V}\int_{0}^{W} \Sigma(E)\left(-\frac{\partial f}{\partial E}\right)dE$. Finite bandwidth of our definition would arise for a band that is abruptly crossed by numerous perfect energy-filtering states acting as perfect resonance levels, or a band that sharply and discontinuously flattens out. Admittedly neither is achievable to perfection in real life, but that theoretical limit is of our interest. We consider an isotropic parabolic band under DPS and optimize $E_{\text{F}}$. Unlike previous studies, we find that there does exist finite, optimum bandwidth for thermoelectrics that depends on temperature and $\kappa_{\text{lat}}$, as delineated by Fig. \ref{fig:bandwidth}.

Achieving $W_{\text{opt}}$ would be a tremendous boost for $zT$. Assuming $\kappa_{\text{lat}}<0.5$ W m$^{-1}$ K$^{-1}$ and $m=0.067$, $W_{\text{opt}}$ elevates $zT$ well beyond 10 - higher than any value attainable through any plain band structures of the previous sections. For given $\kappa_{\text{lat}}$, $W_{\text{opt}}$ generally increases with temperature, as expected from the larger range of carrier excitation at higher temperatures. This implies that achieving $W_{\text{opt}}$ is particularly consequential for low temperatures ($T\le300$ K) where a difference of $0.1\sim0.2$ eV can force a shift in $zT$ by nearly an order of magnitude. As $\kappa_{\text{lat}}$ vanishes, $W_{\text{opt}}$ also vanishes, and $zT$ diverges. This would be the Mahan-Sofo limit, named after their seminal work that deduced widthless band to be optimal, if $\kappa_{\text{lat}}=0$ \cite{bestthermoelectric}. Our recovery of this limit is also evidenced by Supplementary Fig. 8f , in which the `electronic-part' $zT$, labeled $z_{e}T$, diverges to infinity as the band completely flattens out. In the other extreme, as $\kappa_{\text{lat}}$ becomes very high, $W_{\text{opt}}$ diverges, i.s., it is virtually irrelevant for $zT$. Further analysis is provided in Supplementary Discussion.

\section{Discussion}

Commonly used descriptors for thermoelectric performance include the quality factor (QF), which under DPS is \cite{pbsequalityfactor,qualityfactor}
\begin{equation}\label{eq:qualityfactor}
\beta=T\frac{2k_{\text{B}}^{2}}{3\pi}\frac{\rho v_{s}^{2}N_{v}}{m\Delta^{2}\kappa_{\text{lat}}},
\end{equation}
and the Fermi surface complexity factor \cite{complexity}
\begin{equation}\label{eq:complexity}
C=N_{v}\left(\frac{2}{3}\left(\frac{m_{\perp}}{m_{\parallel}}\right)^{-\frac{1}{3}}+\frac{1}{3}\left(\frac{m_{\perp}}{m_{\parallel}}\right)^{\frac{2}{3}}\right)^{3/2}.
\end{equation}
Both metrics promote small effective mass ($m$ or $m_{\parallel}$) and high band multiplicity ($N_{v}$); the latter further promotes band anisotropy $\left(\frac{m_{\perp}}{m_{\parallel}}\right)$. This study serves as a general assessment of these well-known blueprints in thermoelectrics, confirming some while offering fresh perspectives and more complete physical pictures to others.

1. Small $m$ in the transport direction is always better.

2. Band anisotropy is very beneficial, but the extent depends on its type. The advantage of anisotropy draws largely from the fact that $\langle v^{2}(E) \rangle$ rises to steeper, low-dimensional slopes. Bidirectional anisotropy mimicking 1D band structure is particularly beneficial, capable of increasing maximum $zT$ by nearly threefold for a given $m$ in the light direction. 

3. Although not captured by Eq. \ref{eq:qualityfactor} or Eq. \ref{eq:complexity}, the gains from pocket multiplicity and band convergence depend on the relative shapes of the bands and what is assumed of interband scattering. A heavier pocket in the presence of a lighter pocket can be detrimental. Metrics such as QF or $C$ always predict better performance in the presence of more bands because they do not possess any component that accounts for inter-scattering or differential intrinsic transport of each bands. The metrics ought take these effects into account by bounding the gain from $N_{v}$, e.g., using a term such as $N_{v}\left(1+s_{\text{int}}(N_{v}-1)\right)^{-1}$ as was previously described.

4. Within the limits of our investigation, the type of scattering mechanism does not play a pivotal role in determining what band structure is optimal for $zT$, except in the context of resonance levels. In other words, the best-performing band structure is for the most part the same under the DPS, POS, or IIS. The type of scattering decides how much $zT$ improves or suffers as a band transforms, but no transformation is decisively beneficial under one scattering regime but decisively detrimental under another. As one exception, resonance levels are beneficial if the dominant scattering mechanism is efficient at energy-filtering - DPS or POS. If an ineffective filtering mechanism, such as IIS, activates comparably or dominates, then resonance levels lose merit.

5. There exist optimum bandwidths to a plain parabolic band if the transport contribution can be, albeit hypothetically, abruptly curtailed at some energy. Optimum bandwidth arises because low-energy states are undesired owing to their large contribution to $\sigma$ and high-energy states are undesired owing to their large contribution to $\kappa_{e}$. $W_{\text{opt}}$ for a given $m$ depends on temperature and $\kappa_{\text{lat}}$. It is small ($< 0.3$ eV) so long as $\kappa_{\text{lat}}<1$ W m$^{-1}$ K$^{-1}$, and can push $zT$ beyond what is normally accessible.

We stress that our investigation of optimum bandwidth has distinct characteristics from previous investigations in terms of both approach and conclusion. Mahan and Sofo deduced that a fully localized, widthless transport distribution (a completely flat band) would deliver maximal thermoelectric performance \cite{bestthermoelectric}, but under the assumption that $\kappa_{\text{lat}}=0$. Because $\kappa_{\text{lat}}>0$ in real materials, a widthless band and transport distribution would yield $zT=0$ as $v(E)$ and the PF vanish alike (see Fig. \ref{fig:3dsingleopt}b Zone 1). In later studies of optimum bandwidth, the `full-width' definition of bandwidth, $E_{\text{max}}-E_{\text{min}}$, was adopted \cite{optimalbandwidth,bestbandstructure}. A major limitation of this set-up is that the full-width is inherently coupled with $m$ or the size of the BZ. Because smaller $m$ in the transport direction is always beneficial, bandwidth optimality must be probed independently under a fixed $m$, as done in this study. Indeed, in Ref. \cite{optimalbandwidth}, it was determined that optimum full-width does not exist (is infinite) under $\tau(E)\propto D^{-1}(E)$ as $zT$ continues to increase with larger full-width, likely due to concomitantly decreasing $m$. In contrast to these studies, we herein find temperature-dependent, finite optimum bandwidth in the presence of finite $\kappa_{\text{lat}}$. Our bandwidth represents scenarios whereby a band flattens abruptly or features such as high-energy resonance levels are engineered. Our conclusions are more practically relevant than vanishing bandwidth under zero $\kappa_{\text{lat}}$ or infinite full-width that is coupled with $m$ or the BZ size. 

6. Though more of a philosophical point, we propose that analysis of thermoelectrics be more frequently framed in terms of the `thermoelectric conductivity,' $\zeta$, which offers more straightforward insights than that framed in terms of the Seebeck coefficient and Ohmic conductivity. By juxtaposing $\zeta$ against $\sigma$ and $\kappa_{e}$, it becomes clear that a band must develop high $\Sigma(E)$ in the mid-energy region to be optimal for thermoelectric application. Our finding of finite optimum bandwidth resonates with this intuition.

Reflection on real materials is also in order. In spite of the theoretically remarkable performance of extremely anisotropic, `flat-and-dispersive band structures,' they in practice would be subject to a disadvantage due to polycrystallinity of commercial-scale materials as well as symmetry considerations. Indeed, no candidate materials thus far not achieved the high $zT$ modeled here, and we project why in light of our modeling. We distinguish bands in a cubic cell from those in a non-cubic cell.

In a non-cubic cell, a flat-and-dispersive band limits light transport to only certain direction(s). Assuming polycrystallinity, conductivity through a series of differently oriented grains is best described by the lower Wiener bound for composite media, i.e., the harmonic average of directional conductivities \cite{heterogeneousmedia}. Due to poor conductivities along the heavy branch(es), the harmonic average seriously hampers the overall performance under our model, as described in Supplementary Fig. 9. An anisotropic band is then never as good as its isotropic counterpart whose polycrystalline-averaged conductivities are identical to those in any principal direction.

Bi$_{2}$PdO$_{4}$ \cite{bi2pdo4prediction} and BaPdS$_{2}$ \cite{bapds2flatdispersive} are good examples as neither compound is cubic but exhibit bidirectional-anisotropic flat-and-dispersive valence bands. The DOS profiles are characterized by peak-like protrusions near the band edges followed by decays, confirming the 1D-like band structure. Polycrystalline Bi$_{2}$PdO$_{4}$ has been experimentally synthesized and investigated, but recorded rather disappointing $p$-type PF (1 mW m$^{-1}$ K$^{-2}$) and $zT$ (0.06) \cite{bi2pdo4experiment}. Because electronic transport is mobile only in one direction and inhibited in the two heavy directions by design, it is unlikely that their presumably high thermoelectric potential in the light direction would shine through unless the sample is a single crystal. BaPdS$_{2}$ has not yet been tested, but it is reasonable to hypothesize that it may exhibit a similar behavior.

Under cubic symmetry, all three principal directions are guaranteed the same number of light and heavy branches. Polycrystallinity may then be irrelevant here, but now the concern is that the coexistence of light and heavy branches in the direction of transport (as opposed to light in the transport direction, heavy in other directions) with inter-scattering between them can be inherently limiting (recall Zone 1 in Fig. \ref{fig:3ddoubleopt}). Relevant cases are Fe-based full-Heuslers \cite{fe2yz} and perovskite SrTiO$_{3}$ \cite{srtio3lowdimensional}, which exhibit unidirectional flat-and-dispersive conduction bands, with the 2D-like, precipitous DOS at the band edge.

Fe$_{2}$TiSi, a member of the former family, is particularly intriguing because its flat-and-dispersive conduction bands are opposed by triply-degenerate, isotropic valence bands, offering a direct comparison of the $n$-type and $p$-type performances of the respective band structures. According to our DFT \cite{dft} calculation using the PBE functional \cite{pbe}, the lowermost conduction band is very flat along $\Gamma-X$ whose energy width is 0.05 eV ($m_{\parallel}\approx 41$) and dispersive in other directions ($m_{\perp}\approx0.7$). A second isotropic conduction band ($m\approx0.4$) is degenerate at $\Gamma$. Opposing them are three isotropic valence bands with comparable $0.4\le m \le 0.75$. Theoretical thermoelectric properties were studied with rigorous first-principles treatment of electron-phonon scattering \cite{ba2biau}, but the $n$-type PF (5 mW m$^{-1}$ K$^{-2}$ at 300 K) was predicted to be only barely higher than the $p$-type PF (4 mW m$^{-1}$ K$^{-2}$ at 300 K), with no sign of the pronounced performance promised by the flat-and-dispersiveness.

As for SrTiO$_{3}$, according to our DFT calculation, the width of the heavy branch of the lowermost conduction band along $\Gamma-X$ is approximately 0.1 eV ($m_{\parallel}\approx 7$) while in the dispersive directions $m_{\perp}\approx 0.8$. Two additional relatively isotropic conduction bands ($0.4\le m \le0.7$) disperse from $\Gamma$ at the CBM. Multiple experimental reports exist for SrTiO$_{3}$ on single crystals, which should be the best-performing and the most comparable to theory results. Although respectable $n$-type PFs of 3.6 mW m$^{-1}$ K$^{-2}$ \cite{srtio3experiment1} and 2.3 mW m$^{-1}$K$^{-2}$ \cite{srtio3experiment2} have been recorded at room temperature, neither are these values anywhere near what Fig. \ref{fig:3dsingleopt} promises.

These observations collectively suggest that cubic symmetry may cap the full potential of the flat-and-dispersive bands in real materials. As a separate point, it would certainly help if the band masses in the light direction of the both compounds were much smaller.

As a final deliberation, we address the question, what then is the optimal band structure, all things considered? The literature has convincing cases for both an extremely anisotropic, flat-and-dispersive band and a band with multiple dispersive pockets at off-symmetry points. Reflecting on our modeling, we conclude the following. For a single band, if a bidirectional flat-and-dispersive band attaining $\frac{m_{\perp}}{m_{\parallel}}\approx1000$ with as small a possible $m_{\parallel}$ can be realized in a single crystal, it would constitute the optimal single band structure. Otherwise, a band with multitude of dispersive pockets at off-symmetry points with weak inter-scattering would be the best targets, as they provide moderate anisotropy and would be more immune to polycrystallinity. For a given single band, presence of additional bands with equally light mass in the transport direction would increases performance, though this benefit is negligible if $\kappa_{e}>\kappa_{\text{lat}}$ and/or if $s_{\text{int}}\approx1$. For a given overall intrinsic band structure, resonance levels and optimum bandwidth further improves performance, the latter being capable of boosting $zT$ to the highest values of all band structure designs considered here and particularly consequential for low-temperature operation.

As efforts to discover and design thermoelectrics with $zT>3$ continue, the blueprints for high performance grow increasingly influential. Common rules regarding beneficial band structures for bulk thermoelectrics are largely drawn from simple band models without realistic scattering. Using a straightforward but improved approach, we herein fine-tune those blueprints while proposing optimal band structures and design principles along the way. Our generalized findings from this modeling study are mutually supportive of and consistent with the findings from recent targeted studies of high-performing materials with high-fidelity first-principles computations \cite{ba2biau,heuslers,cosiyi,mos2yi,analoguepbte}. We hope the theoretical investigations of the present study help the community navigate rationally towards next-generation thermoelectrics.

\section{Methods}

The Hartree atomic units ($\hbar=m_{e}=a_{o}=q=4\pi\epsilon_{0}=1$) are used throughout the methods section. All calculations are performed on a set of in-house Mathematica codes.

\subsection{Band Structure} 

To generate a realistic solid-state band structure, a band is created by smoothly connecting an upward paraboloid to an inverted paraboloid at a selected inflection point. The advantages of such a band structure include: 1) it remains faithful to solid-state band theory that requires a band to cross the zone-boundary orthogonally, save for when crystal and orbital symmetries allow band crossing or degeneracy at the zone boundary (e.g. in graphene); 2) it formally maintains the validity of effective-mass-based description of the band throughout; 3) relatively simple analytic models of scattering can be directly applied to the upward-parabolic portion, and can be applied with modification for the inverted-parabolic portion; and 4) it can be used to explore a wide range of band structure shapes by modulating the points of inflection. The equation for this band structure is
\begin{widetext}
\begin{equation}\label{eq:band}
\begin{aligned}
E=E_{0}&+\left(\frac{\text{Min}(|k_{x}|,\tilde{k_{x}})}{2m_{\text{up},x}}+\frac{\tilde{k_{x}}^{2}-(G_{x}-\text{Max}(|k_{x}|,\tilde{k_{x}}))^{2}}{2m_{\text{down},x}}\right)
+\left(\frac{\text{Min}(|k_{y}|,\tilde{k_{y}})}{2m_{\text{up},y}}+\frac{\tilde{k_{y}}^{2}-(G_{y}-\text{Max}(|k_{y}|,\tilde{k_{y}}))^{2}}{2m_{\text{down},y}}\right) \\
&+\left(\frac{\text{Min}(|k_{z}|,\tilde{k_{z}})}{2m_{\text{up},z}}+\frac{\tilde{k_{z}}^{2}-(G_{z}-\text{Max}(|k_{z}|,\tilde{k_{z}}))^{2}}{2m_{\text{down},z}}\right) \\
&\pm \left(\frac{|\tilde{k_{x}}^{2}-(G_{x}-\tilde{k_{x}}^{2})^{2}|}{2m_{\text{down},x}}+\frac{|\tilde{k_{y}}^{2}-(G_{y}-\tilde{k_{y}}^{2})^{2}|}{2m_{\text{down},y}}+\frac{|\tilde{k_{z}}^{2}-(G_{z}-\tilde{k_{z}}^{2})^{2}|}{2m_{\text{down},z}}\right).
\end{aligned}
\end{equation}
\end{widetext}
Here, $\tilde{k_{x}}$ denotes the inflection point in the $x$-direction, and $G_{x}$ is the reciprocal lattice vector in the $x$-direction, i.e., the BZ boundary in the $x$-direction. If inflection occurs at halfway to the BZ boundary, then $\tilde{k_{x}}=G_{x}/2$. The last terms are subtracted ($-$) if $\tilde{k_{x}}\ge G_{x}/2$, and added ($+$) if $\tilde{k_{x}}<G_{x}/2$. The same is true in the $y$ and the $z$ directions. Effective masses of the inverted parabolic portion ($m_{\text{down}}$) are obtained by enforcing derivative continuity at the inflection point in every direction.

Although a broad range of band shapes could be explored by changing the inflection point $\tilde{k}$ in any three Cartesian directions, in this work we limit ourselves to bands that inflect halfway to the zone boundary in all three directions. Under this assumption, the effective mass (inverse curvature) profiles of the upward paraboloid portion and the inverted paraboloid portion are identical ($m_{\text{up}}=m_{\text{down}}$), rendering the entire band structure describable with one common set of directional effective masses. We note that this band could also serve as a first-order approximation of the tight-binding cosine band. We create these model bands centered at $\Gamma$ in a simple cubic Brillouin zone corresponding to an arbitrary lattice parameter of $15$ $a_{o}$ (Bohr radius) $\sim 7.9$ \r{A}, which is a reasonable lattice parameter for a real thermoelectric. The density of states (DOS) is calculated using the tetrahedron method \cite{tetrahedron}. Detailed diagrams of the band structure and DOS are given in Supplementary Figs. 1 and S2.

\subsection{Carrier Scattering and Transport} 

Thermoelectric properties are computed by numerically integrating Eqs. \ref{eq:sigma}--\ref{eq:kappae}. The BZ is sampled to convergence with a \textbf{k}-point mesh of $40\times40\times40$.  We fix the effective mass of the principal band in the transport direction ($m_{x}$), unless it evolves isotropically, to enable fair comparison of performance of various band structures. We also ignore bipolar transport except in the two-band case with a conduction and a valence band. This is roughly tantamount to assuming a band gap larger than 0.4 eV - the maximum energy range of thermal excitation when the Fermi level is placed at the band minimum.

The ultimate objective is to determine band structures that theoretically maximize thermoelectric performance, and for that, some settings in place are those known to be beneficial for thermoelectrics. For instance, we intentionally fix $\kappa_{\text{lat}}$ to a low value of 0.5 W m$^{-1}$ K$^{-1}$ as is the case in many phonon-glass materials \cite{tetrahedrite,lonepair1,snsenature,clathrateapl,tl3vse4science}. We fix the principal band mass in the transport direction ($m_{x}$) to a small value of 0.05, in the range of GaAs (0.067) and InSb (0.014). Band anisotropy is also taken to the extreme to explore the limits of its benefits. These sometimes lead to prediction of higher $zT$ ($\sim 9$) than commonly encountered in the literature. Though these settings are optimistic and difficult to simultaneously satisfy experimentally, they are far from unrealistic, as materials with lower $m$ or $\kappa_{\text{lat}}$ are known. They constitute the right regime of exploration in the discourse of high-performance thermoelectrics, and provide some estimate of the realistic upper limit of bulk thermoelectric performance. 

To calculate $\tau$, various scattering mechanisms, namely deformation-potential scattering (DPS) by (acoustic) phonons, polar-optical scattering (POS), and ionized-impurity scattering (IIS), are treated according to well-established formalisms \cite{ziman,nolassharpgoldsmid,lundstrom,frolich,brooks,ionizedimpurity} with appropriate adjustments to account for anisotropy, inverted parabolicity and the BZ-bounded nature of our band structures, with details to follow in the next subsections. Once the lifetimes under the three mechanisms are calculated, the overall $\tau$ is estimated by Matthiesen's rule \cite{matthiessen}.

Supplementary Fig. 3  plots the scattering profiles of the three processes based on Eqs. \ref{eq:defpottau}, \ref{eq:postaunew}, and \ref{eq:brooksherringnew}. For the isotropic cases, they exhibit excellent agreement with the corresponding parabolic band formulations. DPS exhibits the expected $\tau^{-1}\propto D(E)$ relation. POS is characterized by the emission onset near the band edge followed by a gradual decay. IIS reproduces the Brooks-Herring behavior.

\subsection{Deformation-Potential Scattering} 

 The theory of deformation potential was originally developed by Bardeen and Shockley for long-wavelength acoustic phonon scattering with the assumption of elasticity \cite{defpotbardeenshockley}. Combining it with the generalized deformation potential by Kahn and Allen \cite{defpotkahnallen}, we write the DPS lifetimes  as follows:
\begin{equation}\label{eq:defpottau}
\tau_{\text{DPS},\mathbf{k}}=\frac{\rho v_{s}^{2}}{\pi k_{\text{B}}T(\Delta+m\mathbf{v_{k}}\cdot\mathbf{v_{k}})^{2}}D^{-1}(E_{\mathbf{k}}),
\end{equation}
where $\Delta$ is the Bardeen-Shockley deformation potential at the band edge (band shift with lattice deformation) and the Kahn-Allen correction term $m\mathbf{v_{k}}\cdot\mathbf{v_{k}}$ accounts for the shift in the reciprocal space vectors with deformation, which grows large near the zone boundary. 

When more than one bands/pockets are present, interband/intervalley scattering must be accounted for.  The matrix elements for inter-scattering are not obtainable without the details of the phonon spectrum and the Hamiltonian, which we lack because we are not studying a real system. Nevertheless, it is reasonable to assume that inter-scattering is generally weaker than intra-scattering because the wavefunction overlap between distinct bands or pockets is generally weaker than that within a band, and inter-scattering is often heavily reliant on zone-boundary phonons, which are less populated than zone-center phonons. To account for the strength of inter-scattering as such, we introduce a parameter, $s_{\text{int}}$, that acts as the lumped effect of the above-mentioned considerations. This quantity modulates the inter-scattering strength relative to intraband scattering.  We set $s_{\text{int}}=0.5$ to reflect the usually weaker inter-scattering, which allows for trivial extension of Eq. \ref{eq:defpottau} by accounting for the added phase space due to the second band proportional to its DOS. The overall DPS lifetime, for band 1, now becomes
\begin{widetext}
\begin{equation}\label{eq:intervalley}
\tau_{\text{DPS},1\mathbf{k}}=\frac{\rho v_{s}^{2}}{\pi k_{\text{B}}T (\Delta+m_{1}\mathbf{v}_{1\mathbf{k}}\cdot\mathbf{v}_{1\mathbf{k}})^{2}} \left[ D_{1}(E_{1\mathbf{k}})+ s_{\text{int}}D_{2}(E_{1\mathbf{k}}) \right]^{-1},
\end{equation}
\end{widetext}
where the subscripts 1 and 2 indicate each of two bands. The addition of inter-scattering phase space is determined by the presence of second band states at given given $E_{1}$. That is, if $D_{2}(E_{1\mathbf{k}})=0$, then there is no inter-scattering.

  In spite of its simplicity, the above formulations work well also for zone-boundary phonon scattering. For instance, Fe$_{2}$TiSi conduction bands are very flat-and-dispersive, allowing significant zone-boundary phonon scattering as well as intervalley/interband scattering. Accurately calculated scattering rates of Fe$_{2}$TiSi using DFT band structures and electron-phonon matrix elements \cite{epw1,epw3} behave essentially as $\tau^{-1}(E)\propto D(E)$ (see Supplementary Fig. 4a) \cite{ba2biau}, which is also the behavior predicted by Eqs. 10-11. The same conclusion holds for  the flat-and-dispersive valence bands of Li$_{2}$TlBi \cite{analoguepbte}. Therefore, the phenomenological treatment by Eqs. \ref{eq:defpottau}--\ref{eq:intervalley} capture the essence of intra- and inter-scattering even in cases with extreme anisotropy.

\subsection{Polar-Optical Scattering} 

We modify the established formula for polar-optical scattering \cite{lundstrom,nolassharpgoldsmid} as to reasonably accounts for band anisotropy and inverted parabolicity:
\begin{widetext}
\begin{equation}\label{eq:postaunew}
\tau_{\text{POS},\mathbf{k}}=\frac{|\mathbf{v_{k}}|}{2\omega_{o}} \left(\frac{1}{\epsilon_{\infty}}-\frac{1}{\epsilon}\right)^{-1} \left[(b(\omega_{o})+1)\cdot\text{sinh}^{-1}\left(\frac{D(E_{\mathbf{k}}-\omega_{o})}{D(\omega_{o})}\right) +  b(\omega_{o})\cdot\text{sinh}^{-1}\left(\frac{D(E_{\mathbf{k}})}{D(\omega_{o})}\right)\right]^{-1},
\end{equation}
\end{widetext}
Modifications come from exchanging $\sqrt{E}$ with our tetrahedron-integrated $D(E)$ and using the \textbf{k}-dependent prior form (instead of the less general $E$-dependent form - see Supplementary Discussion). Use of the group velocity norm ($|\mathbf{v_{k}}|$) obviates the dependence on $m$ and $E$ through the relation $|\mathbf{v_{k}}|=\sqrt{\frac{2E}{m}}$ for an isotropic parabolic band. The optical phonon frequency is represented by $\omega_{o}$, and $b(\omega_{o})$ is the Bose-Einstein population. The left term in the square brackets accounts for phonon emission whereas the right term accounts for phonon absorption. For our band structures, Eq. \ref{eq:postaunew} is exact in the upward-parabolic portions up to the inflection point, and past it, approximates the true lifetimes. When more than one band pocket exists at one \textbf{k}-point, nothing prohibits interband POS from occuring. To account for this, we take the same approach as we do with DPS and use $D(E)=D_{1}(E)+s_{\text{int}}D_{2}(E)$ to enlarge the phase space.

\subsection{Ionized-Impurity Scattering} 

We use a modified version of the Brooks-Herring formula \cite{brooks,ionizedimpurity} that reasonably account for band anisotropy and inverted parabolicity:
\begin{equation}\label{eq:brooksherringnew}
\tau_{\text{IIS},\mathbf{k}}=\frac{\epsilon^{2}|\mathbf{k}|^{4}}{2\pi^{3}N_{i}Z^{2}}D^{-1}(E_{\mathbf{k}})\left(\text{log}(1+\gamma_{\mathbf{k}})-\frac{\gamma_{\mathbf{k}}}{1+\gamma_{\mathbf{k}}}\right)^{-1},
\end{equation}
where the screening term is 
\begin{equation}\label{eq:gammanew}
\gamma_{\mathbf{k}}=\frac{4|\mathbf{k}|^{2}\epsilon k_{\text{B}}T}{n}\left(\frac{F_{\frac{1}{2}}(E_{\text{F}})}{F_{-\frac{1}{2}}(E_{\text{F}})}\right).
\end{equation}
Modification comes from using the \textbf{k}-dependent prior form of the Brooks-Herring formula (instead of its typical energy-dependent form - see Supplementary Methods) and replacing the terms in it that represent the parabolic DOS with our tetrahedron-integrated $D(E)$. For our band structures, this corrected formula is again exact for the upward-parabolic portions from $\Gamma$ to the inflection point, and past it, closely approximates the true lifetimes. When more than one band pocket exists at the same \textbf{k}-point, interband IIS can take place. To account for this, we again use $D(E)=D_{1}(E_{1\mathbf{k}})+s_{\text{int}}D_{2}(E_{1\mathbf{k}})$ to modify the phase space. We assume that one impurity donates or accepts one charge, meaning the effective impurity charge of $Z=1$. This choice forces that the carrier concentration ($n$) effectively equals the impurity concentration ($n=N_{i}$) at appreciable doping levels.

\subsection{Materials parameters} 

To calculate specific values of scattering, material-dependent quantities that control the relative strength of the three scattering mechanisms must be chosen. These include the deformation potential, the dielectric constants, and the optical phonon frequency among others. We select plausible values for these quantities that occur prevalently in real materials, as listed in Table \ref{tab:quantities}. In emphasis, the choice of these values renders the relative strength of each scattering channel  arbitrary. What is not arbitrary is the characteristic thermoelectric behavior of bands under a given scattering regime.

\begin{table}
\begin{tabular}{|ccc|}
\hline
\textbf{Arbitrary Quantity}        & \textbf{Symbol}     & \textbf{Value}   \\ \hline
Temperature                        & $T$                 & 500 K            \\
Density                            & $\rho$              & 5000 kg/m$^{3}$  \\
Sound Velocity                     & $v_{s}$             & 4000 m/s         \\
Deformation Potential              & $\Delta$            & 0.4 Ha = 10.8 eV \\
Interband Scattering Strength      & $s_{\text{int}}$    & $0.5$          \\
Static Dielectric Constant         & $\epsilon$          & 30               \\
High-freq. Dielectric Constant     & $\epsilon_{\infty}$ & 25               \\
Optical Phonon Frequency           & $\omega_{o}$        & $k_{\text{B}}T/2$  \\
Effective Charge of an Impurity    & $Z$                 & 1                \\
Impurity Concentration             & $N_{i}$             & $n/Z$                \\
Lattice Thermal Conductivity       & $\kappa_{\text{lat}}$             & 0.5 Wm$^{-1}$K$^{-1}$
\\\hline
\end{tabular}
\caption{Arbitrary quantities used in the scattering and transport models.} 
\label{tab:quantities}
\end{table}

\section{Data Availability}

The data can be either reproduced or generated using any user-desired parameters using publicly available Mathematica notebooks (see below).

\section{Code Availability}

The Mathematica notebooks are publicly available on the following link: https://github.com/jsyony37/bandmodel.

\section{Acknowledgments}

This work was led by fundings from U.S. Department of Energy, Office of Basic Energy Sciences, Early Career Research Program, which supported J. P. and A. J. Lawrence Berkeley National Laboratory is funded by the Department of Energy under award DE-AC02-05CH11231. V. O. acknowledges financial support from the National Science Foundation Grant DMR-1611507. This work used resources of the National Energy Research Scientific Computing Center, a Department of Energy Office of Science User Facility supported by the Office of Science of the U.S. Department of Energy under Contract No. DE-AC02-05CH11231. J. P. thanks Younghak Kwon of UCLA Mathematics for helpful discussions.

\bibliography{references}

\begin{thebibliography}{10}

\bibitem{complex}
Snyder, J., Toberer, E.
\newblock Complex thermoelectric materials.
\newblock {\em Nat. Mater.} {\bf 7}, 2 (2008).

\bibitem{newandold}
Sootsman, J.~R., Chung, D., Kanatzidis, M.
\newblock New and old concepts in thermoelectric materials.
\newblock {\em Angew. Chem.} {\bf 48}, 46 (2009).

\bibitem{intuition}
Zeier, W.~G. et~al.
\newblock Thinking Like a Chemist: Intuition in Thermoelectric Materials.
\newblock {\em Angew. Chem.} {\bf 55}, 24 (2016).

\bibitem{perspectivesonthermoelectrics}
Zebarjadi, M., Esfarjani, K., Dresselhaus, M.~S., Ren, Z.~F., Chen, G.
\newblock Perspectives on thermoelectrics: from fundamental to device
  applications.
\newblock {\em Energy Environ. Sci.} {\bf 5}:5147 (2012).

\bibitem{thermoelectricmaterials}
Zhang, X., Zhao, L.-D.
\newblock Thermoelectric materials: Energy conversion between heat and
  electricity.
\newblock {\em J. Materiomics.} {\bf 1}, 2 (2015).

\bibitem{compromisesynergy}
Zhu, T. et~al.
\newblock Compromise and Synergy in High-Efficiency Thermoelectric Materials.
\newblock {\em Adv. Mater.} {\bf 29}, 14 (2017).

\bibitem{advancesinthermoelectrics}
Mao, J. et~al.
\newblock Advances in Thermoelectrics.
\newblock {\em Adv. in Phys.} {\bf 67}, 2 (2018).

\bibitem{advancesinthermoelectricmaterials}
He, J., Tritt, T.~M.
\newblock Advances in Thermoelectric Materials Research: Looking back and
  moving forward.
\newblock {\em Science} {\bf 357}, 6358 (2017).

\bibitem{onthetuning}
Yang, J. et~al.
\newblock On the tuning of electrical and thermal transport in thermoelectrics:
  an integrated theory-experiment perspective.
\newblock {\em Npj Comput. Mater.} {\bf 2}, 15015 (2016).

\bibitem{computationalthermoelectrics}
Gorai, P., Stevanovi{\'c}, V., Toberer, E.~S.
\newblock Computationally guided discovery of thermoelectric materials.
\newblock {\em Nat. Rev. Mater.} {\bf 2}, 17053 (2017).

\bibitem{computationalenergymaterials}
Jain, A., Shin, Y., Persson, K.~A.
\newblock Computational predictions of energy materials using density
  functional theory.
\newblock {\em Nat. Rev. Mater.} {\bf 1}, 15004 (2016).

\bibitem{wangsnyderbook}
Wang, H., Pei, Y., LaLonde, A.~D., Snyder, G.~J.
\newblock {\em Material Design Considerations Based on Thermoelectric Quality
  Factor}, pages 3--32.
\newblock Springer, Berlin, Heidelberg, (2013).

\bibitem{valleytronics1}
Xin, J. et~al.
\newblock Valleytronics in Thermoelectric Materials.
\newblock {\em Quantum Mater.} {\bf 3}, 9 (2018).

\bibitem{valleytronics2}
Norouzzadeh, P., Vashaee, D.
\newblock Classication of Valleytronics in Thermoelectricity.
\newblock {\em Sci. Rep.} {\bf 6}, 22724 (2016).

\bibitem{roleofscattering}
Witkoske, E., Wang, X., Lundstrom, M., Askarpour, V., Maassen, J.
\newblock Thermoelectric band engineering: The role of carrier scattering.
\newblock {\em J. Appl. Phys.} {\bf 122}, 17 (2017).

\bibitem{bandalignmentscattering}
Kumarasinghe, C., Neophytou, N.
\newblock Band alignment and scattering considerations for enhancing the
  thermoelectric power factor of complex materials: The case of Co-based
  half-Heusler alloys.
\newblock {\em Phys. Rev. B} {\bf 99}, 19 (2019).

\bibitem{simplescattering}
Maassen, C. R.~J.
\newblock Analysis of simple scattering models on the thermoelectric
  performance of analytical electron dispersions.
\newblock {\em J. Appl. Phys.} {\bf 127}, 6 (2020).

\bibitem{optimalbandwidth}
Zhou, J., Yang, R., Chen, G., , Dresselhaus, M.~S.
\newblock Optimal Bandwidth for High Efficiency Thermoelectrics.
\newblock {\em Phys. Rev. Lett.} {\bf 108}:226601 (2011).

\bibitem{bestbandstructure}
Jeong, C., Kim, R., Lundstrom, M.~S.
\newblock On the best bandstructure for thermoelectric performance: A Landauer
  perspective.
\newblock {\em J. Appl. Phys.} {\bf 111}:113707 (2012).

\bibitem{onsager1}
Onsager, L.
\newblock Reciprocal Relations in Irreversible Processes I.
\newblock {\em Phys. Rev.} {\bf 37}:405--426 (1931).

\bibitem{onsager2}
Onsager, L.
\newblock Reciprocal Relations in Irreversible Processes II.
\newblock {\em Phys. Rev.} {\bf 38}:2265--2279 (1931).

\bibitem{callen}
Callen, H.
\newblock The Application of Onsager's Reciprocal Relations to Thermoelectric,
  Thermomagnetic, and Galvanomagnetic Effects.
\newblock {\em Phys. Rev.} {\bf 73}, 11 (1948).

\bibitem{dovertheoretical}
Jones, W., March, N.~H.
\newblock {\em Theoretical Solid State Physics}.
\newblock Dover Books (1985).

\bibitem{ziman}
Ziman, J.~M.
\newblock {\em Electrons and Phonons: the theory of transport phenomena in
  solids}.
\newblock Oxford University Press (1960).

\bibitem{btecoefficients}
Scheidemantel, T.~J., Ambrosch-Draxl, C., Thonhauser, T., Badding, J.~V., Sofo,
  J.~O.
\newblock Transport coefficients from first-principles calculations.
\newblock {\em Phys. Rev. B.} {\bf 68}:125210 (2003).

\bibitem{boltztrap}
Madsen, G.~K., Singh, D.~J.
\newblock BoltzTraP. A code for calculating band-structure dependent
  quantities.
\newblock {\em Comput. Phys. Commun.} {\bf 175}:67--71 (2006).

\bibitem{loweffmass}
Pei, Y., LaLonde, A.~D., Wang, H., Snyder, G.~J.
\newblock Low Effective Mass Leading to High Thermoelectric Performance.
\newblock {\em Energy Environ. Sci.} {\bf 5}, 7 (2012).

\bibitem{materialdescriptors}
Yan, J. et~al.
\newblock Material descriptors for predicting thermoelectric performance.
\newblock {\em Energy Environ. Sci.} {\bf 8}, 3 (2015).

\bibitem{complexity}
Gibbs, Z.~M. et~al.
\newblock Effective mass and Fermi surface complexity factor from ab initio
  band structure calculations.
\newblock {\em Npj Comput. Mater.} {\bf 3}, 8 (2017).

\bibitem{halfheuslerbanddegeneracy}
Fu, C. et~al.
\newblock High Band Degeneracy Contributes to High Thermoelectric Performance
  in $p$-type Half-Heusler Compounds.
\newblock {\em Adv. Energy Mater.} {\bf 4}:1400600 (2014).

\bibitem{zintlorbitalengineering}
Zhang, J. et~al.
\newblock Designing high-performance layered thermoelectric materials through
  orbital engineering.
\newblock {\em Nat. Commun.} {\bf 7}, 10892 (2016).

\bibitem{chalcopyritebandconvergence1}
Zhang, J. et~al.
\newblock High-performance pseudocubic thermoelectric materials from non-cubic
  chalcopyrite compounds.
\newblock {\em Adv. Mater.} {\bf 26}, 23 (2014).

\bibitem{bandconvergencereview}
Pei, Y., Wang, H., Snyder, G.~J.
\newblock Band engineering of thermoelectric materials.
\newblock {\em Adv. Mater.} {\bf 24}, 46 (2012).

\bibitem{pbtebandconvergence}
Pei, Y., Xiaoya, S., Aaron, L., H.~Wang, L.~C., Snyder, G.
\newblock Convergence of electronic bands for high performance bulk
  thermoelectrics.
\newblock {\em Nature} {\bf 473}, 7345 (2011).

\bibitem{bandconvergence}
Park, J. et~al.
\newblock When Band Convergence is Not Beneficial for Thermoelectrics.
\newblock {\em arXiv} , 2012.02272 (2020).

\bibitem{cosiyi}
Xia, Y., Park, J., Zhou, F., Ozoli\c{n}\v{s}, V.
\newblock High Thermoelectric Power Factor in Intermetallic CoSi Arising from
  Energy Filtering of Electrons by Phonon Scattering.
\newblock {\em Phys. Rev. Appl.} {\bf 11}, 2 (2019).

\bibitem{cosisame}
Pshenay-Severin, D.~A., Ivanov, Y.~V., Burkov, A.~T.
\newblock The effect of energy-dependent electron scattering on thermoelectric
  transport in novel topological semimetal CoSi.
\newblock {\em J. Phys.: Condens. Matter} {\bf 30}, 47 (2018).

\bibitem{mos2yi}
Xia, Y., Park, J., Ozoli\c{n}\v{s}, V., Wolverton, C.
\newblock Leveraging Electron-Phonon Interaction to Enhance Thermoelectric
  Power Factor in Graphene-Like Semimetals.
\newblock {\em Phys. Rev. B (R)} {\bf 100}:201401 (2019).

\bibitem{ybal3epw}
Lianga, J., Fana, D., Jianga, P., Liua, H., Zhao, W.
\newblock First-principles study of the thermoelectric properties of
  intermetallic compound YbAl$_{3}$.
\newblock {\em Intermetallics} {\bf 87}:27--30 (2017).

\bibitem{thermoelectricsemimetals}
Markov, M., Rezaei, S.~E., Sadeghi, S.~N., Esfarjani, K., Zebarjadi, M.
\newblock Thermoelectric properties of semimetals.
\newblock {\em Phys. Rev. Mater.} {\bf 3}, 9 (2019).

\bibitem{asymmetricbands}
Markov, M., Rezaei, S.~E., Sadeghi, S.~N., Esfarjani, K., Zebarjadi, M.
\newblock Ultra-High Thermoelectric Power Factors in Narrow Gap Materials with
  Asymmetric Bands.
\newblock {\em J. Phys. Chem. C} {\bf 124}, 33 (2020).

\bibitem{bestthermoelectric}
Mahan, G.~D., Sofo, J.
\newblock The Best Thermoelectric.
\newblock {\em Proc. Natl. Acad. Sci.} {\bf 93}:7436--7439 (1996).

\bibitem{pbsequalityfactor}
Wang, H., Pei, Y., LaLonde, A.~D., Snyder, G.~J.
\newblock Weak electron-phonon coupling contributing to high thermoelectric
  performance in $n$-type PbSe.
\newblock {\em Proc. Nat. Acad. Sci.} {\bf 109}, 25 (2012).

\bibitem{qualityfactor}
Zhang, X. et~al.
\newblock Electronic quality factor for thermoelectrics.
\newblock {\em Sci. Adv.} {\bf 6}, 46 (2020).

\bibitem{heterogeneousmedia}
Torquato, S.
\newblock {\em Random Heterogeneous Materials}.
\newblock Springer (2002).

\bibitem{bi2pdo4prediction}
He, J. et~al.
\newblock Bi$_{2}$PdO$_{4}$: A Promising Thermoelectric Oxide with High Power
  Factor and Low Lattice Thermal Conductivity.
\newblock {\em Chem. Mater.} {\bf 29}, 6 (2016).

\bibitem{bapds2flatdispersive}
Isaacs, E., Wolverton, C.
\newblock Remarkable thermoelectric performance in BaPdS$_{2}$ via pudding-mold
  band structure, band convergence, and ultralow lattice thermal conductivity.
\newblock {\em Phys. Rev. Mater.} {\bf 3}, 1 (2019).

\bibitem{bi2pdo4experiment}
Kayser, P., Serrano-Sanchez, F., Dura, O.~J., Fauth, F., Alonso, J.~A.
\newblock Experimental corroboration of the thermoelectric performance of
  Bi$_{2}$PdO$_{4}$ oxide and Pb-doped derivatives.
\newblock {\em J. Mater. Chem. C} {\bf 8}, 16 (2020).

\bibitem{fe2yz}
Bilc, D.~I., Hautier, G., Waroquiers, D., Rignanese, G.-M., Ghosez, P.
\newblock Low-Dimensional Transport and Large Thermoelectric Power Factors in
  Bulk Semiconductors by Band Engineering of Highly Directional Electronic
  States.
\newblock {\em Phys. Rev. Lett.} {\bf 114}, 13 (2015).

\bibitem{srtio3lowdimensional}
Dylla, M.~T., Kang, S.~D., Snyder, G.~J.
\newblock Effect of Two-Dimensional Crystal Orbitals on Fermi Surfaces and
  Electron Transport in Three-Dimensional Perovskite Oxides.
\newblock {\em Angew. Chem.} {\bf 131}, 17 (2019).

\bibitem{dft}
Kohn, W., Sham, L.~J.
\newblock Self-consistent equations including exchange and correlation effects.
\newblock {\em Phys. Rev.} {\bf 140}, 4A (1965).

\bibitem{pbe}
Perdew, J.~P., Burke, K., Ernzerhof, M.
\newblock Generalized gradient approximation made simple.
\newblock {\em Phys. Rev. Lett.} {\bf 77}, 18 (1996).

\bibitem{ba2biau}
Park, J., Xia, Y., Ozoli\c{n}\v{s}, V.
\newblock High Thermoelectric Power Factor and Efficiency from a Highly
  Dispersive Band in Ba$_{2}$BiAu.
\newblock {\em Phys. Rev. Appl.} {\bf 11}, 1 (2019).

\bibitem{srtio3experiment1}
Okuda, T., Nakanishi, K., Miyasaka, S., Tokura, Y.
\newblock Large thermoelectric response of metallic perovskites:
  Sr$_{1-x}$La$_{x}$TiO$_{3}$ ($0\le x\le0.1$).
\newblock {\em Phys. Rev. B} {\bf 63}, 11 (2001).

\bibitem{srtio3experiment2}
Okuda, T., Nakanishi, K., Miyasaka, S., Tokura, Y.
\newblock High-temperature carrier transport and thermoelectric properties of
  heavily La- or Nb-doped SrTiO$_{3}$ single crystals.
\newblock {\em J. Appl. Phys.} {\bf 97}, 3 (2005).

\bibitem{heuslers}
Park, J., Xia, Y., Ganose, A., Jain, A., Ozoli\c{n}\v{s}, V.
\newblock High Thermoelectric Performance and Defect Energetics of
  Multipocketed Full Heusler Compounds.
\newblock {\em Phys. Rev. Appl.} {\bf 14}, 2 (2020).

\bibitem{analoguepbte}
He, J., Xia, Y., Naghavi, S.~S., Ozoli\c{n}\v{s}, V., Wolverton, C.
\newblock Designing chemical analogs to PbTe with intrinsic high band
  degeneracy and low lattice thermal conductivity.
\newblock {\em Nat. Commun.} {\bf 10}, 719 (2019).

\bibitem{tetrahedron}
Bl$\ddot{\text{o}}$chl, P.~E., Jepsen, O., Anderson, O.
\newblock Improved tetrahedron method for Brillouin-zone integrations.
\newblock {\em Phys. Rev. B} {\bf 49}, 23 (1994).

\bibitem{tetrahedrite}
Lu, X. et~al.
\newblock High performance thermoelectricity in earth-abundant compounds based
  on natural mineral tetrahedrites.
\newblock {\em Adv. Energy Mater.} {\bf 3}, 3 (2013).

\bibitem{lonepair1}
Nielsen, M.~D., Ozoli\c{n}\v{s}, V., Heremans, J.
\newblock Lone pair electrons minimize lattice thermal conductivity.
\newblock {\em Energy Environ. Sci.} {\bf 6}:570--578 (2013).

\bibitem{snsenature}
Zhao, L. et~al.
\newblock Ultralow thermal conductivity and high thermoelectric figure of merit
  in SnSe crystals.
\newblock {\em Nature} {\bf 508}:373--377 (2014).

\bibitem{clathrateapl}
Nolas, G.~S., Cohn, J.~L., Slack, G.~A., Scuhjman, S.~B.
\newblock Semiconducting Ge clathrates: Promising candidates for thermoelectric
  applications.
\newblock {\em Appl. Phys. Lett.} {\bf 73}, 2 (1998).

\bibitem{tl3vse4science}
Mukhopadhyay, S. et~al.
\newblock Two-channel model for ultralow thermal conductivity of crystalline
  Tl$_{3}$VSe$_{4}$.
\newblock {\em Science} {\bf 360}, 6396 (2018).

\bibitem{nolassharpgoldsmid}
Nolas, G.~S., Sharp, J., Goldsmid, H.~J.
\newblock {\em Thermoelectrics}.
\newblock Springer (2001).

\bibitem{lundstrom}
Lundstrom, M.
\newblock {\em Fundamentals of Carrier Transport}.
\newblock Cambridge University Press (2000).

\bibitem{frolich}
Fr{\"o}hlich, H.
\newblock Electrons in lattice fields.
\newblock {\em Adv. in Phys.} {\bf 3}:325--361 (1954).

\bibitem{brooks}
Brooks, H.
\newblock Theory of the Electrical Properties of Germanium and Silicon.
\newblock {\em Adv. Elec. Elec. Phys.} {\bf 7}:85--182 (1955).

\bibitem{ionizedimpurity}
Chattopadhyay, D., Queisser, H.~J.
\newblock Electron Scattering by Ionized Impurities in Semiconductors.
\newblock {\em Rev. Mod. Phys.} {\bf 53}, 4 (1981).

\bibitem{matthiessen}
Matthiessen, A.
\newblock Ueber die elektrische Leitungsf{\"a}higkeit der Legirungen.
\newblock {\em Ann. Phys.} {\bf 186}, 6 (1860).

\bibitem{defpotbardeenshockley}
Bardeen, J., Shockley, W.
\newblock Deformation Potentials and Mobilities in Non-Polar Crystals.
\newblock {\em Phys. Rev.} {\bf 80}, 1 (1950).

\bibitem{defpotkahnallen}
Khan, F.~S., Allen, P.~B.
\newblock Deformation potentials and electron-phonon scattering: Two new
  theorems.
\newblock {\em Phys. Rev. B} {\bf 29}, 6 (1984).

\bibitem{epw1}
Giustino, F., Cohen, M.~L., Louie, S.~G.
\newblock Electron-phonon interaction using Wannier functions.
\newblock {\em Phys. Rev. B} {\bf 76}, 16 (2007).

\bibitem{epw3}
Ponce, S., Margine, E.~R., Verdi, C., Giustino, F.
\newblock EPW: Electron--phonon coupling, transport and superconducting
  properties using maximally localized Wannier functions.
\newblock {\em Comput. Phys. Commun.} {\bf 55}:116--133 (2016).

\end{thebibliography}


\begin{thebibliography}{10}

\bibitem{ba2biau}
Park, J., Xia, Y., Ozoli\c{n}\v{s}, V.
\newblock High Thermoelectric Power Factor and Efficiency from a Highly
  Dispersive Band in Ba$_{2}$BiAu.
\newblock {\em Phys. Rev. Appl.} {\bf 11}, 1 (2019).

\bibitem{lowdimensional3d}
Parker, D., Chen, X., Singh, D.~J.
\newblock High Three-Dimensional Thermoelectric Performance from
  Low-Dimensional Bands.
\newblock {\em Phys. Rev. Lett.} {\bf 119}, 14 (2013).

\bibitem{quantumwell}
Hicks, L.~D., Dresselhaus, M.
\newblock Effect of quantum-well structures on the thermoelectric Figure of
  merit.
\newblock {\em Phys. Rev. B} {\bf 47}, 19 (1993).

\bibitem{resonancelevelreview}
Heremans, J.~P., Wiendlochaac, B., Chamoire, A.~M.
\newblock Resonant levels in bulk thermoelectric semiconductors.
\newblock {\em Energy Environ. Sci.} {\bf 5}:5510--5530 (2012).

\bibitem{bestthermoelectric}
Mahan, G.~D., Sofo, J.
\newblock The Best Thermoelectric.
\newblock {\em Proc. Natl. Acad. Sci.} {\bf 93}:7436--7439 (1996).

\bibitem{optimalbandwidth}
Zhou, J., Yang, R., Chen, G., , Dresselhaus, M.~S.
\newblock Optimal Bandwidth for High Efficiency Thermoelectrics.
\newblock {\em Phys. Rev. Lett.} {\bf 108}:226601 (2011).

\bibitem{defpotbardeenshockley}
Bardeen, J., Shockley, W.
\newblock Deformation Potentials and Mobilities in Non-Polar Crystals.
\newblock {\em Phys. Rev.} {\bf 80}, 1 (1950).

\bibitem{defpotkahnallen}
Khan, F.~S., Allen, P.~B.
\newblock Deformation potentials and electron-phonon scattering: Two new
  theorems.
\newblock {\em Phys. Rev. B} {\bf 29}, 6 (1984).

\bibitem{epw1}
Giustino, F., Cohen, M.~L., Louie, S.~G.
\newblock Electron-phonon interaction using Wannier functions.
\newblock {\em Phys. Rev. B} {\bf 76}, 16 (2007).

\bibitem{epw3}
Ponce, S., Margine, E.~R., Verdi, C., Giustino, F.
\newblock EPW: Electron--phonon coupling, transport and superconducting
  properties using maximally localized Wannier functions.
\newblock {\em Comput. Phys. Commun.} {\bf 55}:116--133 (2016).

\bibitem{analoguepbte}
He, J., Xia, Y., Naghavi, S.~S., Ozoli\c{n}\v{s}, V., Wolverton, C.
\newblock Designing chemical analogs to PbTe with intrinsic high band
  degeneracy and low lattice thermal conductivity.
\newblock {\em Nat. Commun.} {\bf 10}, 719 (2019).

\bibitem{lundstrom}
Lundstrom, M.
\newblock {\em Fundamentals of Carrier Transport}.
\newblock (Cambridge University Press 2000).

\bibitem{nolassharpgoldsmid}
Nolas, G.~S., Sharp, J., Goldsmid, H.~J.
\newblock {\em Thermoelectrics}.
\newblock (Springer 2001).

\bibitem{brooks}
Brooks, H.
\newblock Theory of the Electrical Properties of Germanium and Silicon.
\newblock {\em Adv. Elec. Elec. Phys.} {\bf 7}:85--182 (1955).

\bibitem{ionizedimpurity}
Chattopadhyay, D., Queisser, H.~J.
\newblock Electron Scattering by Ionized Impurities in Semiconductors.
\newblock {\em Rev. Mod. Phys.} {\bf 53}, 4 (1981).

\end{thebibliography}

\end{document}


\title{Supplementary Information: Optimal Band Structure for Thermoelectrics with Realistic Scattering and Bands}
	\author{Junsoo Park}
	\affiliation{\lbnl}
	\author{Yi Xia}
	\affiliation{\northwestern}
	\author{Vidvuds Ozoli\c{n}\v{s}}
	\affiliation{\yale} 	
	\affiliation{\esi} 	
	\author{Anubhav Jain}
	\affiliation{\lbnl}
	\date{\today} 
	\maketitle

\section{Supplementary Figures}

\begin{figure}[hp]
\centering
\includegraphics[width=1 \linewidth]{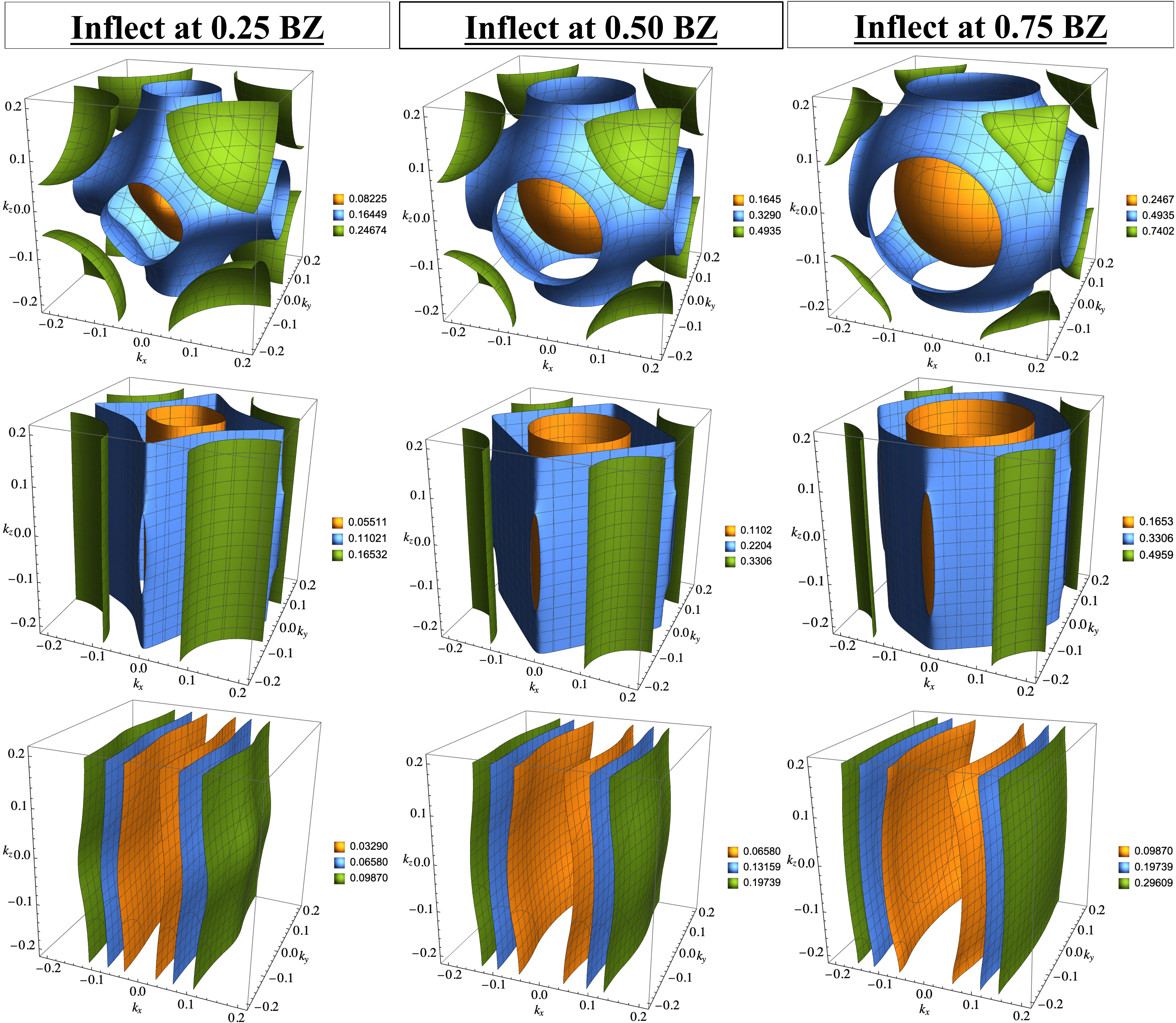}
\caption{Contour plots of the paraboloid + inverted paraboloid band structure with inflection points at 0.25, 0.5, 0.75 way to the BZ boundaries. In the top row, $m_{x}=m_{y}=m_{z}=0.05$ (isotropic). In the middle row, $m_{x}=m_{y}=0.05$ and $m_{z}=5$. In the bottom row, $m_{x}=0.05$ and $m_{y}=m_{z}=5$. The energy scale is in Hartree.}
\label{fig:pinvp}
\end{figure}

\newpage 

\begin{figure}[hp]
\centering
\includegraphics[width=1 \linewidth]{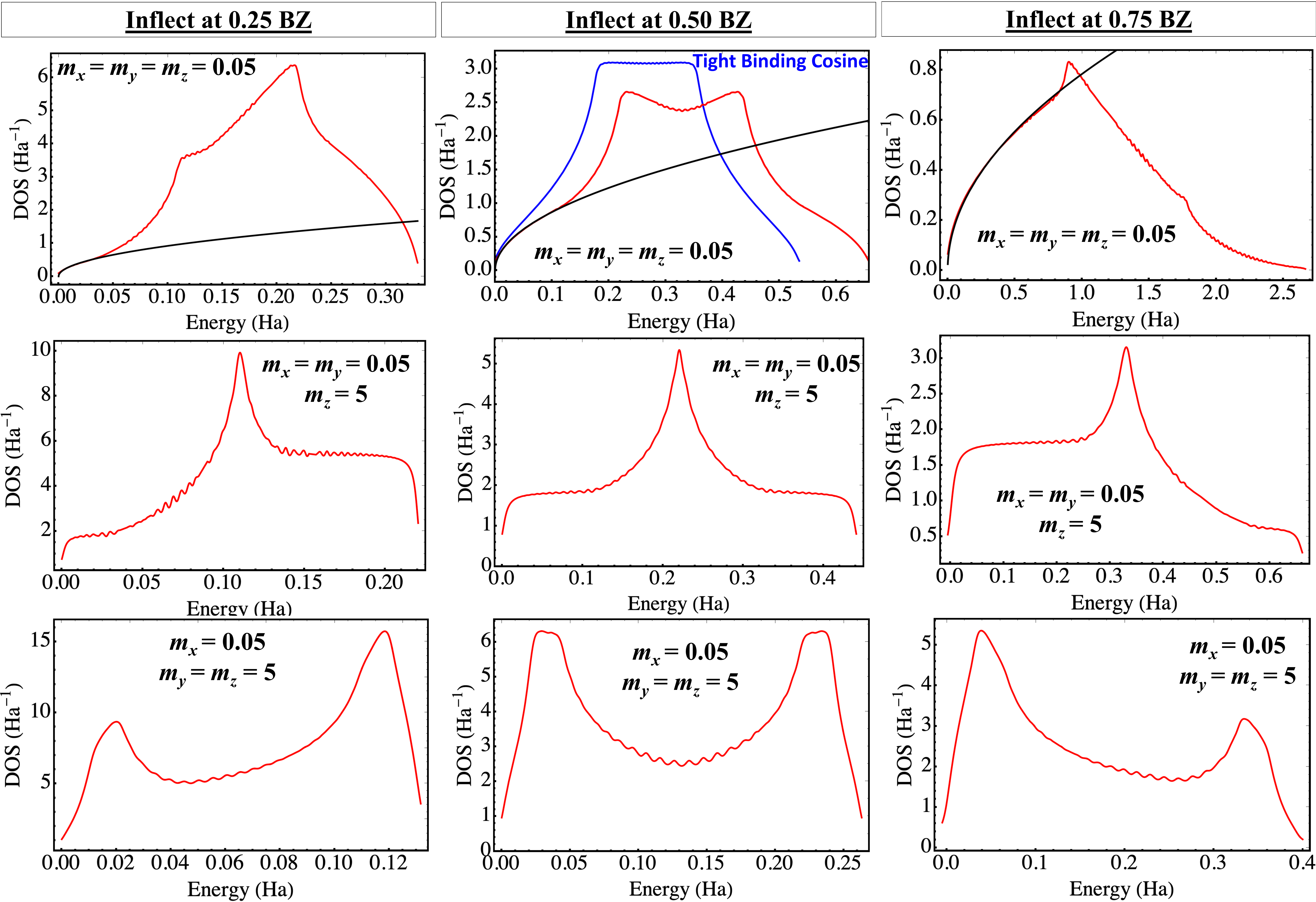}
\caption{The density of states (\color{red}red\color{black}) of paraboloid + inverted paraboloid band structure with inflection points at 0.25, 0.5, 0.75 way to the BZ boundaries. In top row, $m_{x}=m_{y}=m_{z}=0.05$ (isotropy). The parabolic DOS of identical effective mass are plotted in black ($\sim\sqrt(E)$), which agree well with the initial upward-parabolic portion. For the isotropic band inflecting at 0.5 BZ, the similar tight-binding cosine DOS is plotted in \color{blue}blue \color{black} for comparison. In middle row, $m_{x}=m_{y}=0.05$ and $m_{z}=5$ (unidirectional anisotropy), and the DOS onsets resemble the two-dimensional precipice.. In bottom row, $m_{x}=0.05$ and $m_{y}=m_{z}=5$ (bidirectional anisotropy), and the DOS onsets resemble the one-dimensional peak. }
\label{fig:pinvpdos}
\end{figure}

\newpage

\begin{figure*}[hp]
\centering
\includegraphics[width=1 \linewidth]{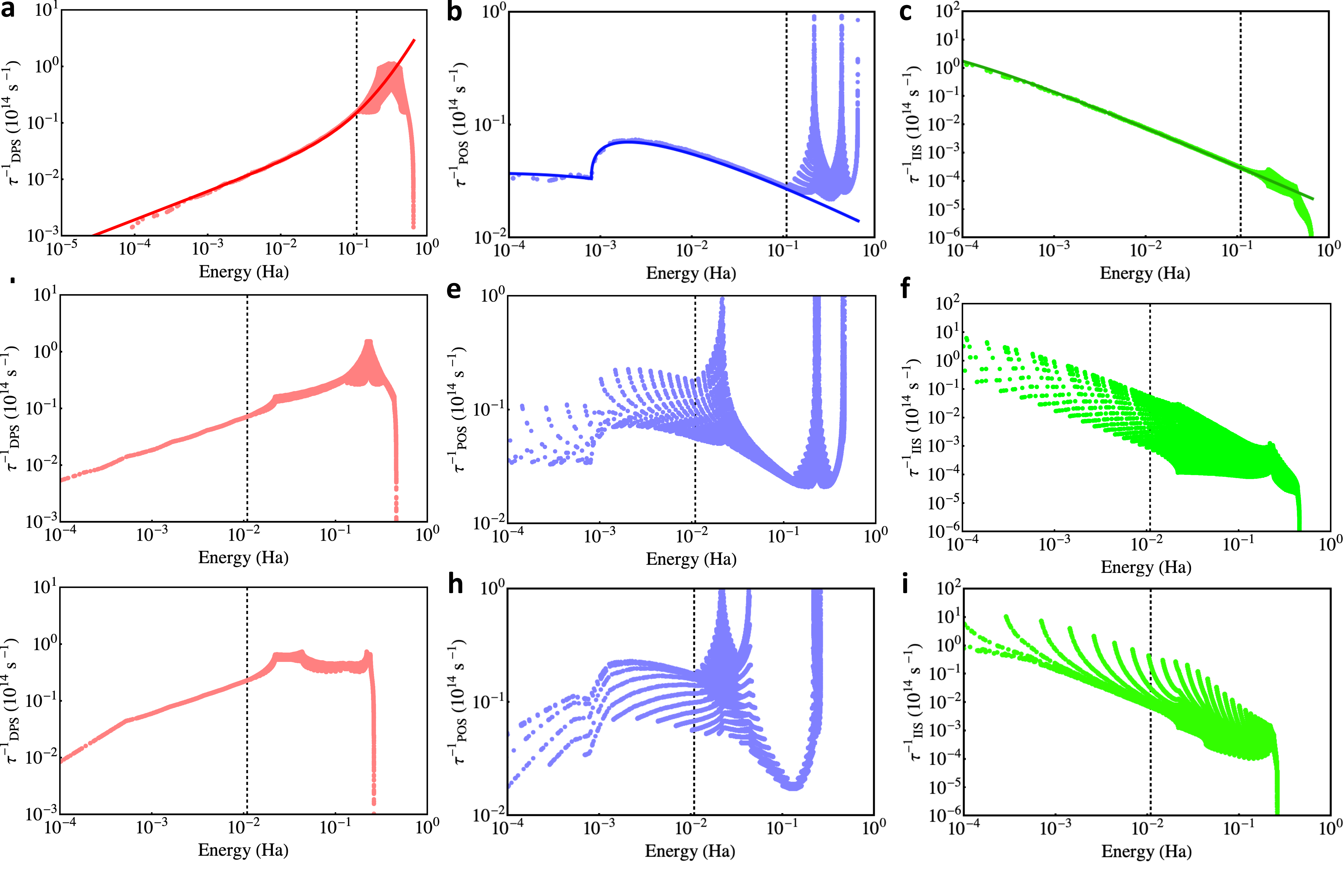}
\caption{The \textbf{k}-dependent scattering rates plotted against energy for the paraboloid + inverted paraboloid band structure inflecting halfway to the zone-boundary. The vertical dashed lines demarcate the end of the purely upward-parabolic regime. \textbf{a, d, g)} Deformation-potential scattering (Eq. 11 in the main text), \textbf{b, e, h)} polar-optical scattering (Eq. 14 in the main text), and \textbf{c, f, i)} ionized-impurity scattering (Eq. 18 in the main text). \textbf{a-c)} Isotropic case, where $m_{x}=m_{y}=m_{z}=0.05$. Overlaid in solid lines are the energy-dependent scattering rates for a generic parabolic band of the same effective mass profile given by Eqs. 11, 13, and 15 in the main text. The agreements are essentially perfect in the upward-parabolic regime (before the dotted vertical lines), followed by deviations due to the downward inflection of our band. \textbf{d-f)} One-way-anisotropic case, where $m_{x}=m_{y}=0.05$ and $m_{z}=0.5$. \textbf{g-i)} Two-way-anisotropic case, where $m_{x}=0.05$ and $m_{y}=m_{z}=0.5$.}
\label{fig:scatter}
\end{figure*}

\newpage

\begin{figure}[hp]
\centering
\includegraphics[width=0.8 \linewidth]{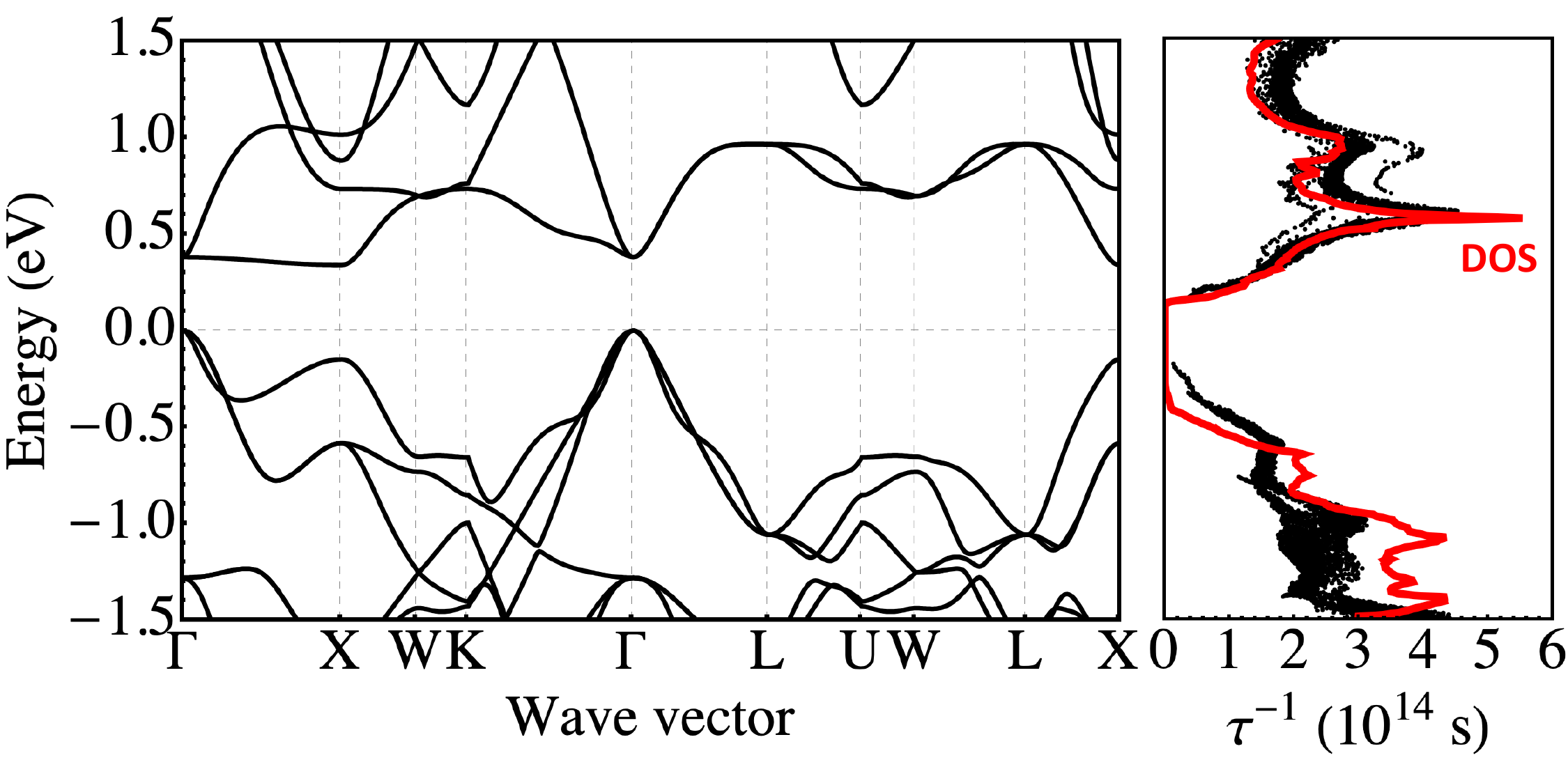}
\caption{The band structure and electron-phonon scattering rates of  Fe$_{2}$TiSi, with DOS overlaid in \color{red}red\color{black}. The scattering rates are calculated with the EPW software. The flat-and-dispersive conduction bands are dominated by DPS, which is essentially proportional to DOS. The valence bands are dominated by POS and does not scale as DOS as well \cite{ba2biau}.} 
\label{fig:scatteringsupp}
\end{figure}

\vspace{5cm}

\begin{figure}[hp]
\centering
\includegraphics[width=0.8\linewidth]{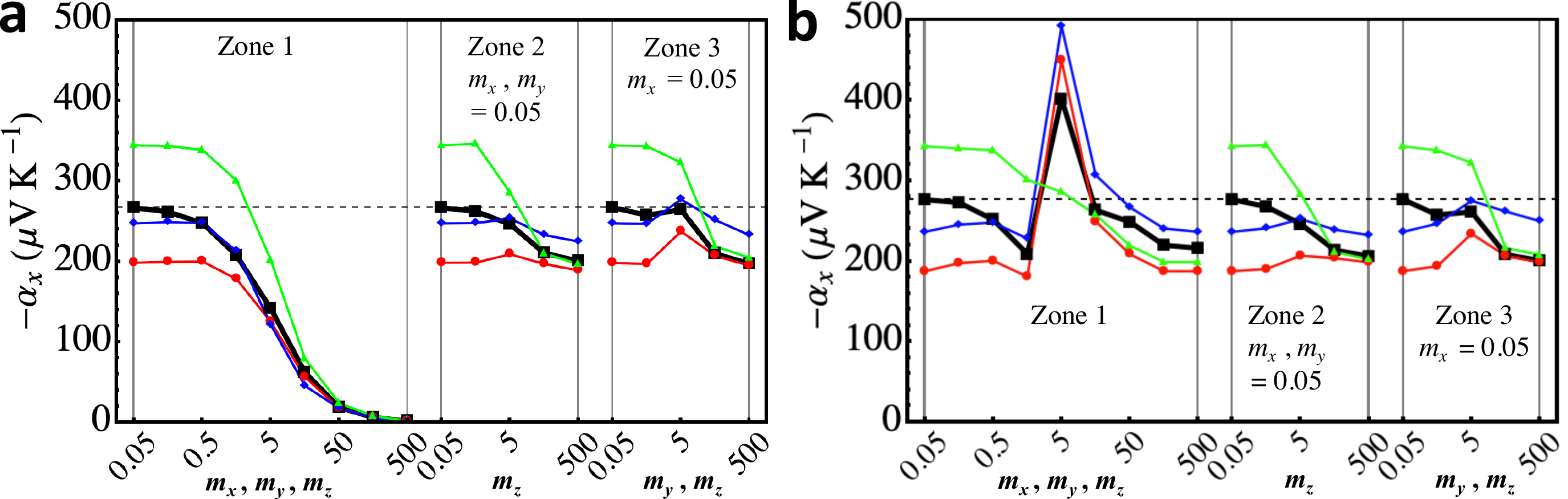}
\caption{The Seebeck coefficient in the light direction ($x$) with the Fermi level fixed to the band minimum, as a function of changing effective masses in three directions. \textbf{a)} Single band evolves as depicted in Fig. 1a of the main text. \textbf{b)} Two bands exist where the second band  evolves as depicted in Fig. 1b of the main text. Each zone indicates certain characteristic evolution: isotropic increase in $m$ from 0.05 to 500 in Zone 1, anisotropic increase in $m_{y}$ from 0.05 to 500 in Zone 2, anisotropic increase in both $m_{y}$ and $m_{z}$ in Zone 3 from 0.05 to 500. Four different scattering regimes are considered: the POS limit (\color{blue}blue\color{black}), the IIS limit (\color{green}green\color{black}), the DPS limit (\color{red}red\color{black}), and the overall effect (black). The black dashed horizontal line marks the isotropic value.} 
\label{fig:3dsingleef0}
\end{figure}

\newpage

\begin{figure}[hp]
\centering
\includegraphics[width=0.85 \linewidth]{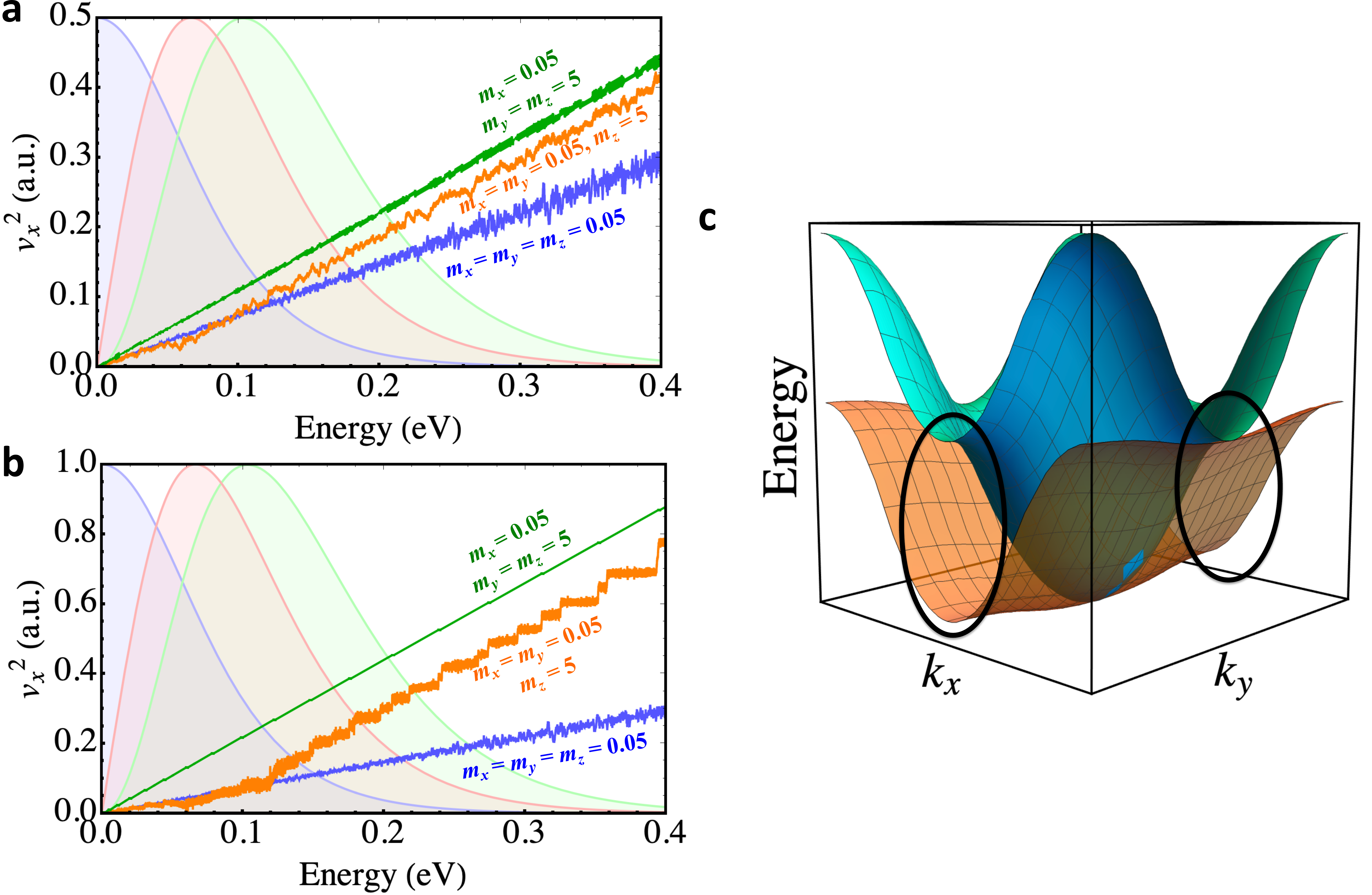}
\caption{Schematics of factors that render band anisotropy's effect on thermoelectric performance beneficial or harmful. \textbf{a)} Evolution of moving-averaged group velocities in the $x$ direction, $\langle v_{x}^{2}(E) \rangle$, for one-way anisotropy. \textbf{b)} Evolution of moving-averaged $\langle v_{x}^{2}(E) \rangle$ for two-way anisotropy. \textbf{c)} Dispersions of two paraboloid + inverted paraboloid bands, one isotropic (blue) and one very anisotropic (orange). Black circles mark the regions ``low-energy voids," where low-energy states exist for the latter but are absent for the former.}
\label{fig:anisotropyprofiles}
\end{figure}

\newpage

\begin{figure}[hp]
\centering
\includegraphics[width=0.8 \linewidth]{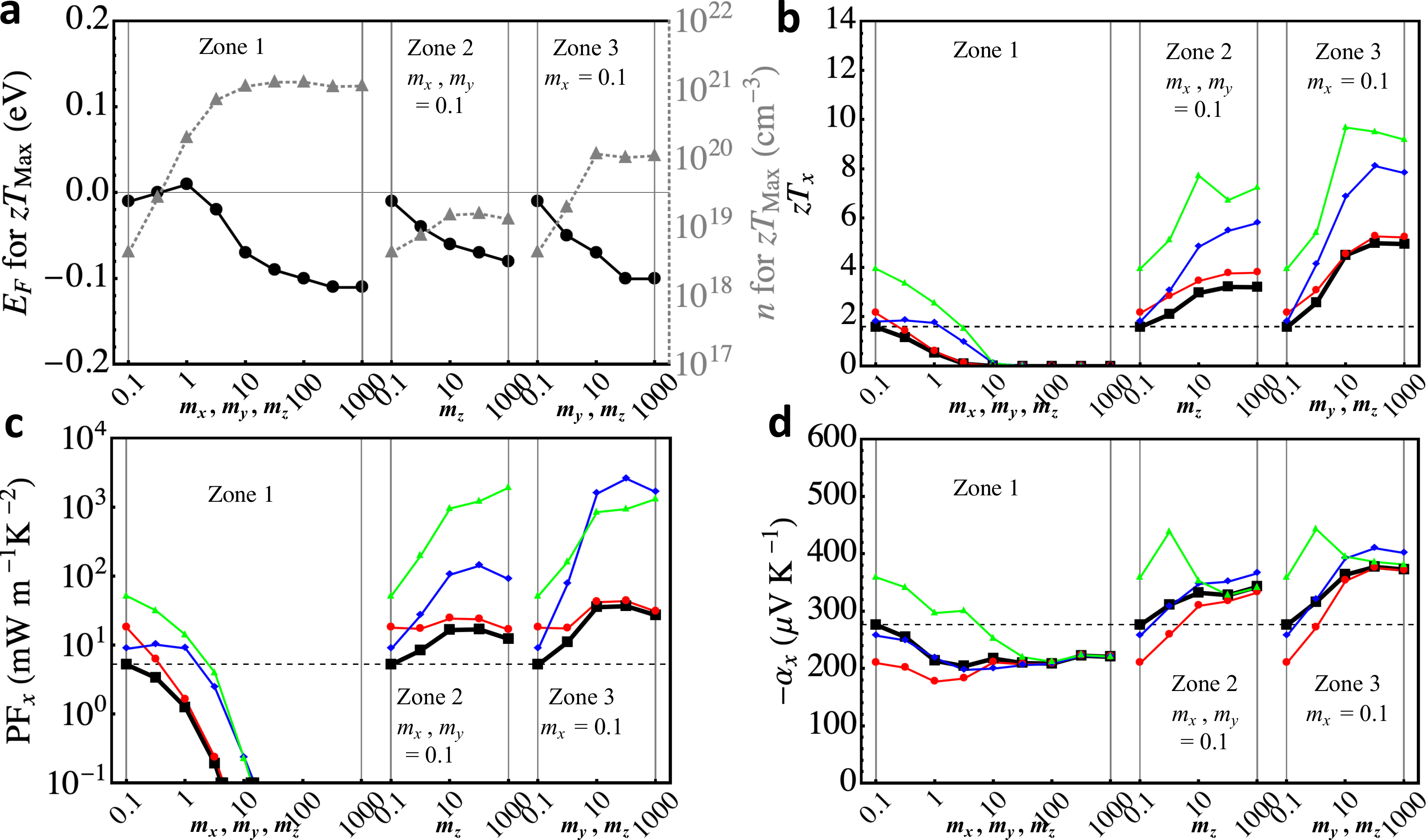}
\caption{Same as Fig. 2 in the main text but with a larger $m_{x}=0.1$ and a higher $\kappa_{\text{lat}}=1$ W m$^{-1}$ K$^{-1}$.}
\label{fig:3dsingleoptlow}
\end{figure}

\vspace{2cm}

\begin{figure}[hp]
\centering
\includegraphics[width=1 \linewidth]{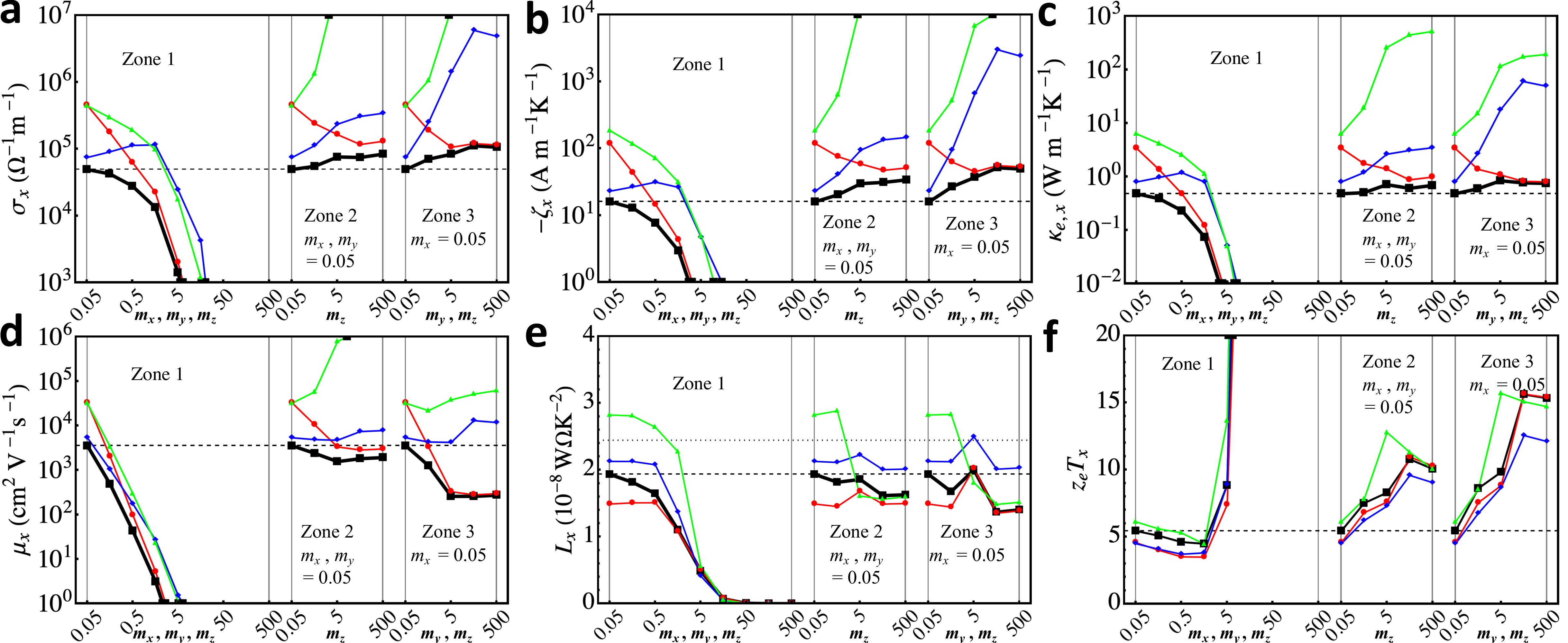}
\caption{Supplement to Fig. 2 in the main text. Electronic transport properties in the light direction ($x$) for the single-band case, where horizontal dashed lines indicate the initial value at $m_{x}=m_{y}=m_{z}=0.15$. \textbf{a)} The charge (Ohmic) conductivities. \textbf{b)} The thermoelectric conductivities. \textbf{c)} The electronic thermal conductivities. \textbf{d)} Mobilities, which in the isotropic region decrease as $\mu_{\text{DPS}}\sim m^{-2.49}$, $\mu_{\text{POS}}\sim m^{-1.49}$, and $\mu_{\text{IIS}}\sim m^{-1.88}$. \textbf{e)} The Lorenz numbers, where the dotted line indicates the Wiedemann-Franz value ($L_{0}=2.44\times10^{-8}$ W$\Omega$K$^{-2}$). \textbf{f)} The electronic-part $zT$.}
\label{fig:3dsingleoptsupp}
\end{figure}

\newpage

\begin{figure}[hp]
\centering
\includegraphics[width=1 \linewidth]{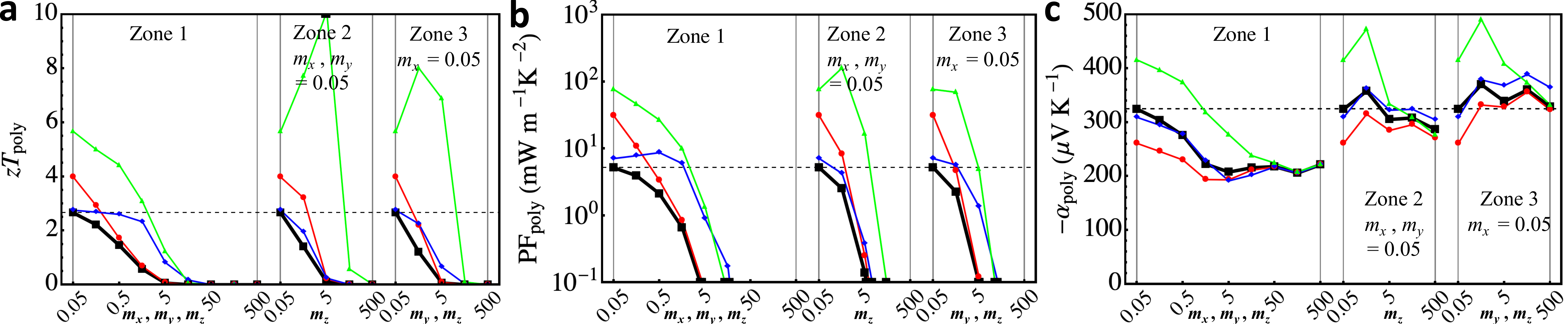}
\caption{Polycrystalline thermoelectric properties of single band approximated by taking the harmonic mean of directional properties, supplemental electronic transport properties to Fig. 3 in the main text. \textbf{a)} $zT$, \textbf{b)} the PF, \textbf{c)} the Seebeck coefficient. The anisotropic zones 2 and 3 are marked by significantly worse performance compared to the single-light-direction performance provided in Fig. 2 in the main text. The isotropic zone 1 is identical to the single-direction trend.}
\label{fig:polycrystalline}
\end{figure}

\vspace{5cm}

\begin{figure}[hp]
\centering
\includegraphics[width=1 \linewidth]{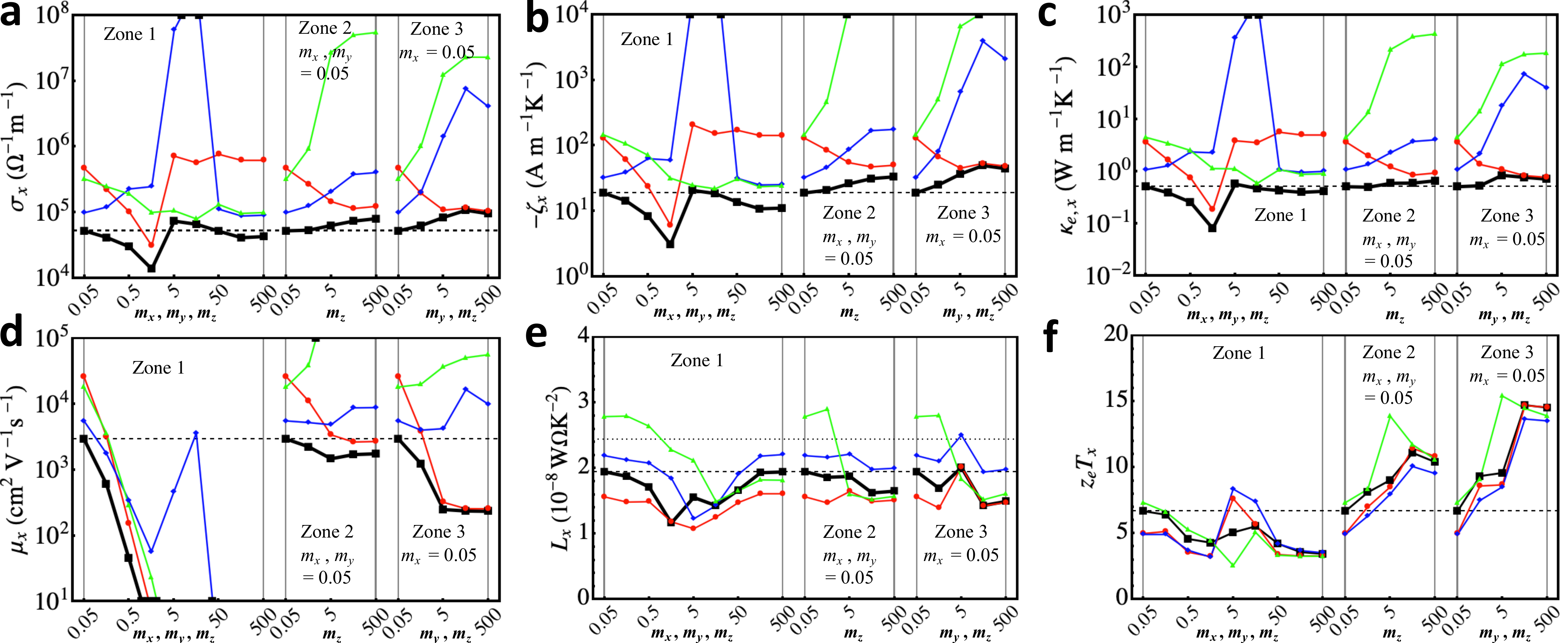}
\caption{Supplement to Fig. 3 in the main text. Electronic transport properties in the light direction ($x$) for the two-band case, where horizontal dashed lines indicate the initial values for the single band case at $m_{x}=m_{y}=m_{z}=0.05$. \textbf{a)} The charge (Ohmic) conductivities. \textbf{b)} The thermoelectric conductivities. \textbf{c)} The electronic thermal conductivities. \textbf{d)} Mobilities, which in the isotropic region decrease as $\mu_{\text{DPS}}\sim m_{2}^{-2.46}$, $\mu_{\text{POS}}\sim m_{2}^{-1.37}$, and $\mu_{\text{IIS}}\sim m_{2}^{-1.81}$ where $m_{2}$ is the effective mass of the evolving second band. \textbf{e)} The Lorenz numbers, where the upper dotted horizontal line indicates the Wiedemann-Franz value ($L_{0}=2.44\times10^{-8}$ W$\Omega$K$^{-2}$). \textbf{f)} The electronic-part $zT$.}
\label{fig:3ddoubleoptsupp}
\end{figure}

\newpage

\begin{figure}[hp]
\centering
\includegraphics[width=0.8 \linewidth]{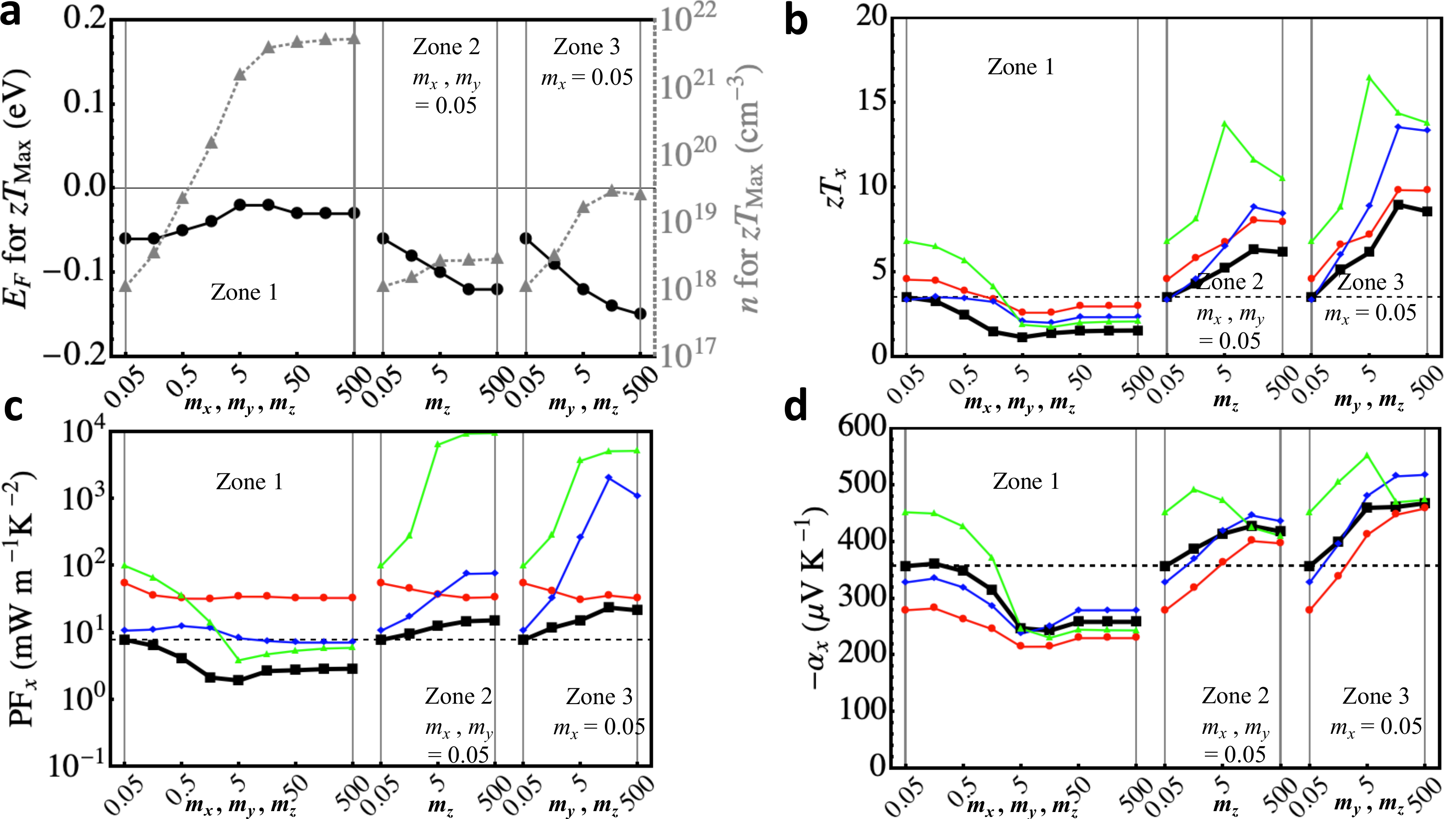}
\caption{Same as Fig. 3 in the main text but with no interband scattering ($s_{\text{int}}=0$).}
\label{fig:3ddoubleoptnoint}
\end{figure}

\vspace{3cm}

\begin{figure}[hp]
\centering
\includegraphics[width=1 \linewidth]{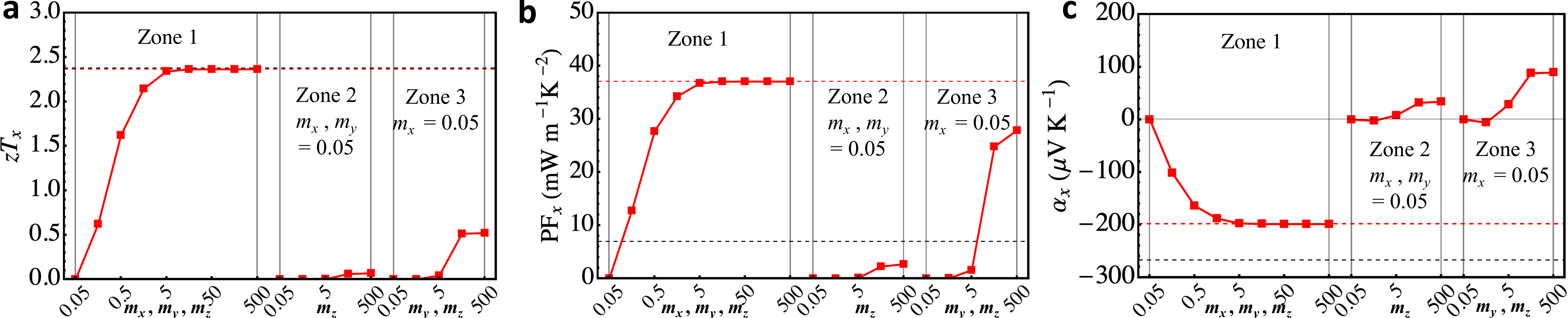}
\caption{Thermoelectric properties in the light direction ($x$) with bipolar effect (zero band gap) as the valence band effective masses evolve and the conduction band is fixed at $m_{x}=m_{y}=m_{z}=0.05$, as in Fig. 1d in the main text. \textbf{a)} The power factor, \textbf{b)} the Seebeck coefficient, and \textbf{c)} $zT$. Each zone indicates certain characteristic evolution of the valence band: isotropic increase in $m$ from 0.05 to 500 in Zone 1, anisotropic increase in $m_{y}$ from 0.05 to 500 in Zone 2, anisotropic increase in both $m_{y}$ and $m_{z}$ in Zone 3 from 0.05 to 500. Only DPS is considered due to the metallic assumption (very high carrier concentrations). The red horizontal dashed lines indicate the single-conduction-band (i.e. insulating state) values under DPS scattering only, and the black horizontal dashed lines indicate those under the overall effect of POP and IIS as well as DPS.} 
\label{fig:3dbipolar}
\end{figure}

\newpage

\begin{figure}[hp]
\includegraphics[width=0.8 \linewidth]{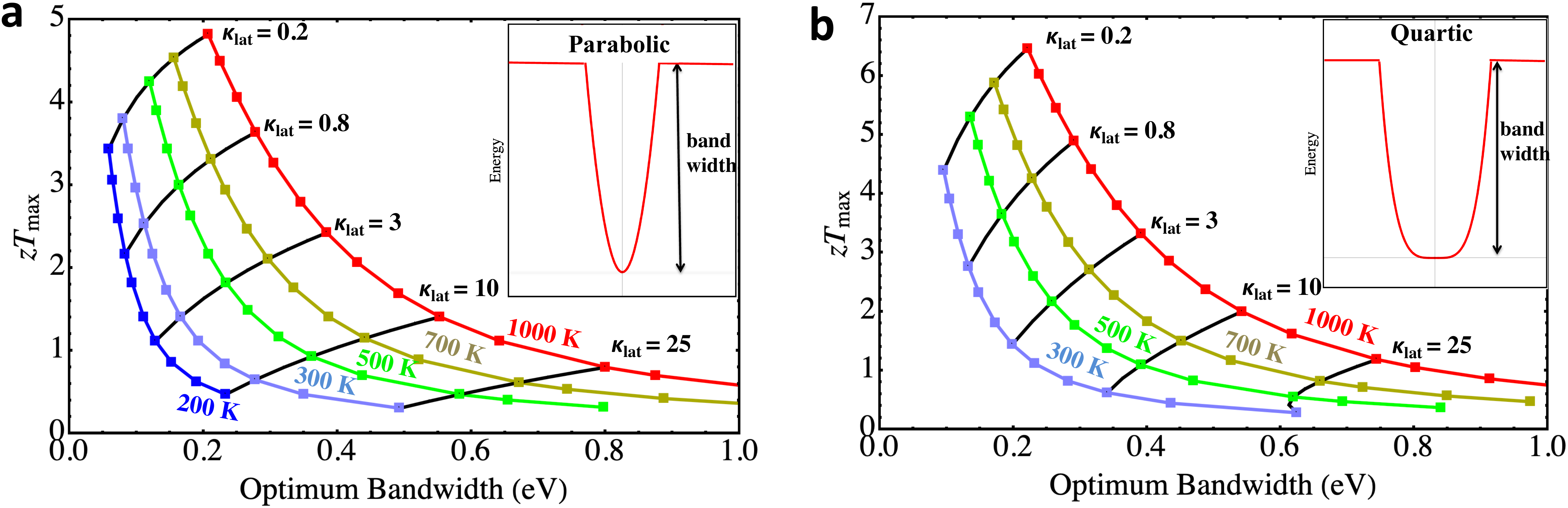}
\caption{$T$-and-$\kappa_{\text{lat}}$-dependent optimum bandwidth and $zT$ under DPS and fixed $E_{\text{F}}=0$ for \textbf{a)} an isotropic 3D parabolic band of $m_{\text{GaAs}}=0.067$, and \textbf{b)} an isotropic 3D quartic with the same energy as the parabolic band at the inflection point. The Fermi level here is fixed at the respective CBM for comparison purposes. $\kappa_{\text{lat}}$ is given in W m$^{-1}$K$^{-1}$.} 
\label{fig:bandwidthef0}
\end{figure}

\newpage

\section{Supplementary Discussion}


A review of the single parabolic band model and the Pisarenko formulas for the Seebeck coefficient is in order owing to their generality, wide reference, and limitations our study improves upon. The Seebeck coefficient is, in the degenerate or metallic limit,
\begin{equation}\label{eq:pisarenko1}
\alpha=\frac{\pi^{2}k_{\text{B}}^{2}T}{3E_{\text{F}}}=\frac{2k_{\text{B}}^{2}}{3}mT\left(\frac{\pi}{3n}\right)^{\frac{2}{3}},
\end{equation}
and in a non-degenerate case,
\begin{equation}\label{eq:pisarenko2}
\begin{aligned}
\alpha&=k_{\text{B}}\left(\frac{5}{2}+s-\frac{E_{\text{F}}}{k_{\text{B}}T}\right) \\
&=k_{\text{B}}\left(\frac{5}{2}+s+\text{log}\left(n^{-1}\left(\frac{mk_{\text{B}}T}{2\pi}\right)^{\frac{3}{2}}\right)\right),
\end{aligned}
\end{equation}
where $n$ is carrier concentration, $m$ is the band effective mass, $k_{\text{B}}$ is the Boltzmann constant, and $s$ is the power of energy to which carrier lifetimes are proportional ($\tau\propto E^{s}$). Contrary to a widely held notion, a heavier band (high $m$) does not by default generate higher Seebeck coefficient under this model since $\alpha$ can be written as a function of either $E_{\text{F}}$ or $m$-and-$n$, where the trivial exchange of variables is allowed by their interrelations:
\begin{equation}\label{eq:fermidegenerate}
E_{\text{F}}=\frac{\pi^{2}}{2m}\left(\frac{3n}{\pi}\right)^{\frac{2}{3}}
\end{equation}
in the degenerate case, and 
\begin{equation}\label{eq:ferminondegenerate}
\frac{E_{\text{F}}}{k_{\text{B}}T }=\text{log}\left(n^{-1}\left(\frac{mk_{\text{B}}T}{2\pi}\right)^{\frac{3}{2}}\right)
\end{equation}
in the non-degenerate case.
Equations~\ref{eq:pisarenko1}--\ref{eq:pisarenko2} stipulate that, if $E_{\text{F}}$ is fixed, then $\alpha$ is constant whatever the $m$ value because $n$ would change accordingly. A light band with a small $m$ would produce the same Seebeck coefficient as a heavier band with a larger $m$ provided that $E_{\text{F}}$ is kept fixed.

The lead-up to Eqs. \ref{eq:pisarenko1}--\ref{eq:pisarenko2} bears one hidden but critical assumption: that there is always enough (infinite) dispersion in all directions to cover the entire energy range relevant to thermoelectric phenomena. However, it breaks down when a band becomes critically heavy that it reaches the BZ boundary before gaining enough energy to cover the entire relevant energy range for transport. Also, because a true solid-state band must at some point change in curvature from positive to negative and cross the BZ boundary orthogonally, there comes a point where assuming constant positive curvature introduces additional unrealistic effects. As will be seen, these effects lead to some conclusions that deviate from what would otherwise be drawn from Eqs. \ref{eq:pisarenko1}--\ref{eq:pisarenko2}. Other assumptions in typical parabolic band models include absence of opposing bands (bipolar effect) and a monotonically behaved, or at least slow-varying, $\Sigma(E)$ such that the Sommerfeld expansion is valid, a requirement for arriving at Eqs. \ref{eq:pisarenko1}--\ref{eq:pisarenko2}.


To demonstrate the importance of our method, we calculate our model-predicted Seebeck coefficient $\alpha$ with a fixed $E_{\text{F}}$ at the band minimum. This exercise helps clarify the difference in behavior of our model versus previous models including SPB, and also clearly illustrates the concept of improving $\zeta$ relative to $\sigma$ via band structure alone. The results are plotted in Supplementary Fig. \ref{fig:3dsingleef0}.

The fixed-$E_{\text{F}}$ results for single band evolution as depicted in Fig. 1a of the main text are given in Supplementary Fig. \ref{fig:3dsingleef0}a. The main takeaway from Zone 1 is the problem of insufficient dispersion for critically heavy bands. We observe that $\alpha$ starts on a plateau that corresponds to the classic SPB model behavior. When the band becomes critically heavy, however, $\alpha$ starts to decrease below the SPB value. The origin of this deviation is that the band becomes heavy enough to terminate at the BZ boundary before gaining enough energy to fully trigger $\zeta$, and misses out on some high-energy states that would otherwise contribute relatively more to $\zeta$ than $\sigma$. For example, at 500 K, carriers of up to 0.18 eV and 0.25 eV above the Fermi level make 97\% of the total contribution to respectively $\sigma$ and $\zeta$. However in our model, if $m=5$, the band encounters the BZ boundary at a value of 0.18 eV, which is enough to almost entirely trigger $\sigma$ but miss important contributions to $\zeta$. Any further increase in $m$ translates to increasingly greater relative loss for $\zeta$ than $\sigma$, continuously degrading the Seebeck coefficient. Typical SPB models overlook this problem of insufficient dispersion owing to the finite-sized BZ, breaking down for heavy enough dispersions. Of note, $\alpha$ goes to 0 as the band completely flattens out, which is explained in the later discussions on optimum bandwidth.

The main lesson from Zones 2 and 3 in Supplementary Fig. \ref{fig:3dsingleef0}a, where the band evolves anisotropically, is the role of group velocity. Under DPS and to a lesser extent under POS, we observe that moderate anisotropy gives the highest $\alpha$, represented by the $\alpha$ peak in the middle of the two zones ($m_{y}, m_{z}=5$). Because $\Sigma(E)\propto v^{2}(E)$ under DPS, thermoelectric behaviors are determined entirely by the average group velocities, $\langle v_{x}^{2}(E) \rangle$. For an isotropic parabolic band, $\langle v^{2}(E) \rangle = \frac{2mE}{3}$; thus, the contribution to $\Sigma(E)$ scales linearly with $E$. For a moderately anisotropic band, $\langle v_{x}^{2}(E) \rangle$ develops a kink. That is, it abruptly steepens in slope (see supplementary Figs. \ref{fig:anisotropyprofiles}a--b). Specifically, for unidirectional anisotropy (heavy only in $z$), $\langle v_{x}^{2}(E) \rangle \propto mE$ post-kink which is the 2D parabolic velocity scaling. For bidirectional anisotropy (heavy in $y$ and $z$), $\langle v_{x}^{2}(E) \rangle \propto 2mE$ post-kink which is the even steeper 1D parabolic velocity scaling. Relative to the scaling of an isotropic band, the kinked $\langle v_{x}^{2}(E) \rangle$ profiles weight $\zeta$ more than $\sigma$ because the velocities increase more steeply at higher energies that at lower energies. This allows $\alpha$ to peak at some moderate anisotropy, and the peak is higher for bidirectional anisotropy. For extremely anisotropic bands mimicking low-dimensional bands, which are also popularly known as ``flat-and-dispersive" bands \cite{lowdimensional3d,quantumwell}, $\langle v_{x}^{2}(E) \rangle$ reverts to linear scaling but with the steeper, post-kink slopes.

Another subtlety regarding extremely anisotropic bands is that they exhaust ``low-energy voids". That is to say, wherever carriers line up along the heavy direction(s), their dispersion in the light direction starts from essentially the band minimum energy. This poses a stark contrast to a less anisotropic band or an isotropic band, for which some dispersion towards the light direction may start from higher energies leaving behind a void of states at lowest energies (see Supplementary Fig. \ref{fig:anisotropyprofiles}c). Because low-energy states contribute relatively more to $\sigma$ than to $\zeta$, their absence is a clear benefit to thermoelectrics. Extremely anisotropic bands exhaust these voids, and therefore, the overall $\alpha$ is somewhat lower than in the isotropic case. POS retains some of the same $\alpha$ signatures of DPS, while under IIS $\alpha$ only decreases with anisotropy.

Next, we examine the case of two bands whose results are in Supplementary Fig. \ref{fig:3dsingleef0}b. Here, we fix the first band in shape and evolve the second band mass according to Fig. 1b in the main text. The results are largely similar to the single-band results but for a prominent peak in $\alpha$ in the middle of Zone 1 (under DPS and POS) where the second band flattens out isotropically. This peak represents the second band acting as a resonance level \cite{resonancelevelreview} that performs energy-filtering due to interband scattering. Although the isotropically heavy band has negligible direct contribution to transport, it can act as a localized scattering partner for the dispersive principal band where their energies overlap, or ``resonate," thereby preferentially scattering low-energy carriers. This increases $\zeta$ relative to $\sigma$ because the low-energy states that had previously contributed much more to $\sigma$ than $\zeta$ are now selectively scattered by the narrowed second band. As a result, $\alpha$ is able to well exceed its single-dispersive-band value. As the second band completely flattens out, however, its width becomes too narrow to filter enough states and hence $\alpha$ is reduced again. We observe that IIS is not a good agent of energy-filtering. 

In summary, whereas $\alpha$ is constant for any band under the SPB model as long as $E_{\text{F}}$ is fixed, our revised model correctly reflects its fluctuating response to changes in a band structure, especially as it approaches extreme shapes.


The optimum $E_{\text{F}}$, plotted in Fig. 2a of the main text, is entirely below the band minimum (zero). This can be understood by noting that generally $\kappa_{e}>>\kappa_{\text{lat}}$ in our model, which in turn leads to $zT\approx\frac{\alpha^{2}}{L}$ where $L$ is the Lorenz number. In the absence of bipolar effect, because $\alpha$ is higher at lower $E_{\text{F}}$ whereas $L$ is relatively constant with respect to $E_{\text{F}}$, $zT$ peaks at low $E_{\text{F}}$ (non-degenerate doping) near where $\alpha$ peaks.

The optimum $E_{\text{F}}$ fluctuates with band evolution. In Zone 1, initially, optimal $E_{\text{F}}$ increases as the band turns heavier. This is because, as the band turns heavier, $\kappa_{e}$ becomes comparable to and then lower than $\kappa_{\text{lat}}$. When $\kappa_{e}<\kappa_{\text{lat}}$, higher PF is required to drive high $zT$. Because the PF is maximized with $E_{\text{F}}$ near the band minimum, optimal $E_{\text{F}}$ increases to meet it. When the band becomes critically heavy and narrow in Zone 1, optimal $E_{\text{F}}$ reverses course and moves away from the band minimum. This is because $E_{\text{F}}$ must again be placed at a distance from the band minimum in order to generate finite $\alpha$ for reasons explained in the later discussions regarding optimum bandwidth.

In Zones 2 and 3, as the band turns anisotropic, $\zeta$ and $\kappa_{e}$ both increase relative to $\sigma$ due to steepening group velocity profile. Because $\kappa_{e}$ increases relative to $\kappa_{\text{lat}}$, which is fixed in our model, $zT$ is increasingly less needy of high PF, and peaks at increasingly lower $E_{\text{F}}$ near where $\alpha$ is maximized.

In Fig. 3a in the main text, describing the multi-band context, the optimal $E_{\text{F}}$ behaves largely similar as it does in Fig. 2a except for the huge spike in the center of Zone 1. A spike in optimal $E_{\text{F}}$, through the band, corresponds to the case where the resonance effect is the most pronounced. Because low-energy states are heavily scattered due to the second band performing energy-filtering, $\zeta$ does not suffer bipolar reduction even if these low-energy states are placed on the opposite side of the Fermi level. Neither does $\sigma$ significantly increase. In turn, $zeta$ benefits from larger $\Sigma(E)$ of the deeper conduction states. When the second band becomes too narrow to act as a resonance level, the optimal $E_{\text{F}}$ again falls below the band minimum as in the single-band case of Fig. 2a.


For a simple study of bipolar effect, we fix the band gap to 0 such that the two bands are tangent to one another at $k=0$ and $E=0$. Note that the band gap is an adjustable parameter in our model. We then fix the dispersion of the conduction band and modulate the valence band effective masses (see Fig. 1c in the main text). Two tangent bands are admittedly not how realistic metallic band structures usually are, but nonetheless, this set-up does probe the essence of how bipolar effect could be resisted for metals and tiny-gap semiconductors. We consider only DPS since the very high $n$ in metals would virtually completely screen Coulombic mechanisms that are POS and IIS.

The message of Supplementary Fig. \ref{fig:3dbipolar} is straightforward: bipolar negation of the Seebeck coefficient is suppressed if one band is light and the opposing band is heavy, or more specifically if there is an asymmetry in $\Sigma(E)$ about the Fermi level. The greater the contrast, the better, though the benefit effectively saturates past a point. From the $n$-type point of view, a completely flat valence band would not carry any hole current but only function as a potential resonance level for low-energy electrons (would require inelastic processes). The desired effect in this picture is essentially a hybrid of the light-band-over-heavy-band rule and energy-filtering: keep holes heavy and filter them further out with as much resonance scattering as possible, while keeping electrons mobile and scattering-free except at very low energies. It is worth recognizing that, in the limit of completely flat valence band, one essentially has a degenerate-semiconducting state with a resonance level and a ``gap" below. This indicates that, within the conventional setting herein assumed, the ideal limit for a metallic thermoelectric is precisely the semiconducting limit with only DPS being present. The identity of the rightmost $zT$ values (under DPS) in each zone in Figs. 3a (main text) and \ref{fig:3dbipolar}a proves the point.

Lastly for metals and tiny-gap semiconductors, $\kappa_{e}$ is frequently far higher than $\kappa_{\text{lat}}$, where small Lorenz number ($L=\frac{\kappa_{e}}{\sigma T}$) becomes critical. Even for typical semiconductors, once $\kappa_{\text{lat}}$ is reduced and the PF is improved, realizing small $L$ would be the final piece of the puzzle. Mirroring the way in which bipolar transport is fought, it is theoretically rather clear what must be done to achieve small $L$: filter out very high-energy states because they contribute relatively more to the thermal current than the thermoelectric or Ohmic currents. They could, in theory, be either 1) filtered with additional states that locally accommodate heavy scattering at high energies, or 2) better yet by simple absence of high-energy states.


Investigation of optimal electronic structures for thermoelectrics can be traced back to the seminal work by Mahan and Sofo \cite{bestthermoelectric}. They took a purely mathematical approach to formulate $zT$ in terms of the energy integrals presented in the main text, and derived the ideal spectral conductivity for maximization of $zT$. They determined that a Dirac delta function near the Fermi level, say at $E_{\dagger}\approx E_{\text{F}}$, is the ideal functional form:
\begin{equation}\label{eq:mahansofo}
\Sigma(E)=D(E)v^{2}(E)\tau(E)=\Sigma_{0}\delta_{E,E^{\dagger}},
\end{equation}
where $\Sigma_{0}$ is some pre-factor. $D(E)$ and $v^{2}(E)$ are directly determined by the electronic structure, while $\tau(E)$ is only indirectly related to it and heavily depends on electron scattering mechanisms. However, their approach was purely mathematical in nature with an implicit assumption that $\kappa_{\text{lat}}=0$. Supplementary Eq. \ref{eq:mahansofo} must be interpreted with caution when applied to reality.

For Supplementary Eq. \ref{eq:mahansofo} to be satisfied, mathematically, at least one of $D(E)$, $v^{2}(E)$, or $\tau(E)$ must be $\delta_{E,E^{\dagger}}$. Physically, however, the terms cannot be independently reduced to a delta function. Firstly, $v^2(E)$ cannot be $\delta_{E,E^{\dagger}}$ while $D(E)$ is not, because $v^2(E)$ is categorically zero without some band dispersion around $E_{\text{F}}$, while no band dispersion can arise at all if $D(E)$ has no width. Secondly, provided there is some dispersion, $\tau(E)$ cannot be a delta function unless electrons are perfectly scattered everywhere but at single $E_{\text{F}}$, which is next to impossible. Then the only plausible way in which $\tau(E)$ can be a delta function is if $D(E)$ is also. These considerations indicate that the only way for Supplementary Eq. \ref{eq:mahansofo} to hold is for $D(E)=N_{v}\delta_{E,E^{\dagger}}$, reflecting one or more ($N_{v}$) perfectly localized states, or perfectly flat bands. The factor $N_{v}$ arises from the fact that eDOS of each band must integrate to 1 (or 2 if spin-degenerate) to conserve the number of electrons:
\begin{equation}\label{eq:edosintegral}
N_{v}=\sum_{1}^{N_{v}}\int_{-\infty}^{\infty} \delta_{E,E^{\dagger}}dE.
\end{equation}
This in turn forces $\tau(E)=\tau_{0}\delta_{E,E^{\dagger}}$ for some finite $\tau_{0}$ limited by elastic scattering, but more importantly it forces $v_{0}\rightarrow0$ and therefore by default $\Sigma(E)\rightarrow0$. 

The implications of $\Sigma(E)\rightarrow0$ are as follows. First, the conductivity is immediately 0, as has also been pointed out by a previous study \cite{optimalbandwidth},
\begin{equation}\label{eq:sigma0}
\sigma=\int_{-\infty}^{\infty} \Sigma(E)\left(-\frac{\partial f}{\partial E}\right)dE=0.
\end{equation}

Second, the Seebeck coefficient is expressible as the following limit as $\Sigma(E)\rightarrow0$ point-by-point:
\begin{equation}\label{eq:mahansofoseebeck1}
\begin{aligned}
\alpha=\lim_{\Sigma(E)\rightarrow0}\frac{\frac{1}{T}\int_{-\infty}^{\infty} \Sigma(E)(E_{\text{F}}-E)\left(-\frac{\partial f}{\partial E}\right)dE}{\int_{-\infty}^{\infty} \Sigma(E)\left(-\frac{\partial f}{\partial E}\right)dE}.
\end{aligned}
\end{equation}
Generally speaking, this is a non-trivial limit to evaluate because $\Sigma(E)$ is a function, not a scalar. However, with the knowledge that $\Sigma(E)$ is widthless (due to $D(E)$), and so it would approach zero at a single point, we can reformulate the limit as
\begin{equation}\label{eq:vsquared}
\lim_{\Sigma(E)\rightarrow0}\Sigma(E)=\lim_{n\rightarrow\infty}\frac{1}{n}\Sigma_{0}\delta_{E,E^{\dagger}},
\end{equation}
where $n\ge1$ is an integer, and evaluate Supplementary Eq. \ref{eq:mahansofoseebeck1} as
\begin{equation}\label{eq:mahansofoseebeck2}
\begin{aligned}
\alpha&=\lim_{n\rightarrow\infty}\frac{\frac{1}{n}\frac{1}{T}\int_{-\infty}^{\infty} \Sigma_{0}\delta_{E,E^{\dagger}}(E_{\text{F}}-E)\left(-\frac{\partial f}{\partial E}\right)dE}{\frac{1}{n}\int_{-\infty}^{\infty} \Sigma_{0}\delta_{E,E^{\dagger}}\left(-\frac{\partial f}{\partial E}\right)dE} \\
&=\frac{1}{T}(E_{\text{F}}-E^{\dagger}),
\end{aligned}
\end{equation}
and this behavior is graphically verified in Fig. 2a and Fig. 3c in the main text. In Fig. 2a, where $E^{\dagger}$ tends to $E_{\text{F}}$ at the band minimum as the band narrows, $\alpha$ tends to 0. In Fig. 3c, where $E^{\text{F}}$ is away from the band minimum that $E^{\dagger}$ tends to, $\alpha$ tends to some finite value corresponding to $(E_{\text{F}}-E^{\dagger})$.

Third, by the same token as above, the Lorenz number can be shown to tend to 0 for a widthless band:
\begin{widetext}
\begin{equation}\label{eq:mahansofolorenz}
\begin{aligned}
L&=\lim_{n\rightarrow\infty}\frac{\frac{1}{n}\left[\frac{1}{T}\int_{-\infty}^{\infty} \Sigma_{0}\delta_{E,E^{\dagger}}(E_{\text{F}}-E)^{2}\left(-\frac{\partial f}{\partial E}\right)dE-T\alpha^{2}\int_{-\infty}^{\infty} \Sigma_{0}\delta_{E,E^{\dagger}}\left(-\frac{\partial f}{\partial E}\right)dE\right]}{\frac{1}{n}\int_{-\infty}^{\infty} \Sigma_{0}\delta_{E,E^{\dagger}}\left(-\frac{\partial f}{\partial E}\right)dE} \\
&=\frac{1}{T}(E_{\text{F}}-E^{\dagger})^{2}-\frac{1}{T}(E_{\text{F}}-E^{\dagger})^{2}=0,
\end{aligned}
\end{equation}
\end{widetext}
and this behavior is graphically suggested in Supplementary Fig. \ref{fig:3dsingleoptsupp}e. By Eqs. \ref{eq:mahansofoseebeck2} and \ref{eq:mahansofolorenz}, given some $E^{\dagger}$, a perfectly localized band of widthless $D(E)$ would lead to divergence in ``electronic-part $zT$," or $zT$ without $\kappa_{\text{lat}}$:
\begin{equation}\label{eq:electronicpartzt}
z_{e}T = \frac{\alpha^{2}\sigma}{\kappa_{e}}T = \frac{\alpha^{2}}{L}=\infty.
\end{equation}
This behavior, consistent with the conclusions of the Mahan-Sofo theory, is graphically suggested in Supplementary Fig. \ref{fig:3dsingleoptsupp}f. However, because of finite Seebeck and vanishing conductivity, the PF would vanish, and compounded by $\kappa_{\text{lat}}>0$ in real materials, $zT$ would vanish to 0 alike. Ergo, even if a set of perfectly localized states could exist in real materials, it is not the physically ideal structure for $zT$ or the PF in real materials. The fundamental barrier is, again, that the components of $\Sigma(E)$ cannot be independently widthless. The value of $z_{e}T$ as a metric for thermoelectric performance improves only as $v(E)$ becomes finite and large and as $\kappa_{\text{lat}}$ is kept minimal.

In supplement to the optimum bandwidth of a parabolic band, we also consider an isotropic quartic band, $E=c(k_{x}^{2}+k_{y}^{2}+k_{z}^{2})^{2}$. The quartic dispersion coefficient $c$ is selected such that it gives the quartic band the same energy as the parabolic band at the inflection point. Supplementary Fig. \ref{fig:bandwidthef0} shows the results. Between parabolic and quartic dispersions, we see that the latter performs better by about 20\%. Since $\tau\propto D^{-1}(E)$ under DPS, it is again the average group velocities distribution that is responsible for the better quartic performance in the transport direction: $\langle v_{x}^{2}(E) \rangle=\frac{16\sqrt{c}}{3}E^{1.5}$. This obviously grows faster with $E$ than that for the parabolic $\langle v_{x}^{2}(E) \rangle=\frac{2}{3m}E$, thereby weighting $\zeta$ relatively more than $\sigma$.

\section{Supplementary Methods}


The deformation-potential scattering was initially developed for long-wavelength acoustic phonon scattering \cite{defpotbardeenshockley}. Even with the Kahn-Allen correction \cite{defpotkahnallen} that we introduce, the overall DPS rate follows the DOS for the most part. However, as mentioned in the main text, the $\tau^{-1}_{\text{DPS}}\propto D(E)$ model works very well in practice for very anisotropic bands in the presence of  zone-boundary phonon scattering as well as interband scattering. This was verified by accurate first-principles calculation of electron-phonon scattering using the EPW software \cite{epw1,epw3} for materials such as Fe$_{2}$TiSi \cite{ba2biau} and Li$_{2}$TlBi \cite{analoguepbte}.

Fe$_{2}$TiSi conduction bands are flat-and-dispersive and dominated by DPS \cite{ba2biau}. The scattering rate for the conduction bands closely follow DOS, which validates the $\tau^{-1}_{\text{DPS}}\propto D(E)$ model. The valence bands are affected significantly by POS, and their scattering rates deviate from DOS. See Supplementary Fig. \ref{fig:scatteringsupp}. Li$_{2}$TlBi valence bands are flat-and-dispersive and dominated by DPS \cite{analoguepbte}. The scattering rate for the valence bands closely follow DOS, which again validates the $\tau^{-1}_{\text{DPS}}\propto D(E)$ model..

Therefore, we expect the extension of Eqs. 10 and 11 in the main text to work well for our model band structures.


For a parabolic band, there exists an analytic formula for energy-dependent lifetime due to inelastic POS \cite{lundstrom,nolassharpgoldsmid},
\begin{widetext}
\begin{equation}\label{eq:postau}
\tau_{\text{POS}}(E)=\frac{E^{\frac{1}{2}}}{\sqrt{2}\omega_{o}m^{\frac{1}{2}}} \left(\frac{1}{\epsilon_{\infty}}-\frac{1}{\epsilon}\right)^{-1} \left[(b(\omega_{o})+1)\cdot\text{sinh}^{-1}\left(\sqrt{\frac{E}{\omega_{o}}-1}\right) +  b(\omega_{o})\cdot\text{sinh}^{-1}\left(\sqrt{\frac{E}{\omega_{o}}}\right)\right]^{-1}.
\end{equation}
\end{widetext}
The first (second) term in the square brackets represents emission (absorption). However, derivation of Supplementary Eq. \ref{eq:postau} uses an energy-momentum phase-space integral simplified with a parabolic dispersion relation, rendering its direct application to non-parabolic bands unjustified. The hyperbolic arcsin terms account for the availability of DOS ($\sim\sqrt{E}$ for a parabolic band) for carriers to be scattered into (final states). Our correction in the main text makes partial corrections by utilizing custom-calculated DOS and \textbf{k}-dependent forms of Supplementary Eq. \ref{eq:postau}.


The established energy-dependent formalism for IIS of a parabolic band is the Brooks-Herring formula \cite{brooks,ionizedimpurity},
\begin{equation}\label{eq:brooksherring}
\tau_{\text{IIS}}(E)=\frac{\sqrt{2}\epsilon^{2}m^{\frac{1}{2}}E^{\frac{3}{2}}}{\pi N_{i}Z^{2}}\left(\text{log}(1+\gamma(E))-\frac{\gamma(E)}{1+\gamma(E)}\right)^{-1},
\end{equation}
where $\gamma$ is a screening term defined as,
\begin{equation}\label{eq:gamma}
\gamma(E)=\frac{8mE\epsilon k_{\text{B}}T}{n}\left(\frac{F_{\frac{1}{2}}(E_{\text{F}})}{F_{-\frac{1}{2}}(E_{\text{F}})}\right).
\end{equation}
and $F_{z}(E_{\text{F}})$ is the Fermi-Dirac integral
\begin{equation}\label{eq:fdintegral}
F_{z}(E_{\text{F}})=\frac{1}{\Gamma(z+1)} \int_{0}^{\infty} \frac{y^{z}}{1+\text{exp}(y-\frac{E_{\text{F}}}{k_{\text{B}}T})}dy.
\end{equation}
Its derivation involves \textbf{k}-space integration over spherical isoenergy surfaces of a parabolic band, rendering it also non-trivial to extend to non-parabolic bands. Our correction in the main text makes partial corrections by utilizing custom-calculated DOS and \textbf{k}-dependent forms of Supplementary Eqs. \ref{eq:brooksherring}--\ref{eq:gamma}.

\section{Supplementary References}

\bibliography{references}